\documentclass[letterpaper,english,aps,prl,reprint,superscriptaddress]{revtex4-1}
\usepackage[latin9]{inputenc}
\usepackage{color}
\usepackage{amsmath}
\usepackage{amssymb}
\usepackage{graphicx}
\usepackage{esint}

\makeatletter

\newcommand*\LyXThinSpace{\,\hspace{0pt}}
\pdfpageheight\paperheight
\pdfpagewidth\paperwidth

\DeclareTextSymbolDefault{\textquotedbl}{T1}

\usepackage{hyperref}
\hypersetup{
    colorlinks = true,
	allcolors = blue	
}
\usepackage{times}

\makeatother

\usepackage{babel}
\begin{document}
\preprint{APS/123-QED}
\title{Surface Chern-Simons theory for third-order topological insulators and superconductors\\
}
\author{Zhi-Hao Huang}
\affiliation{International Center for Quantum Materials and School of Physics, Peking University,
Beijing 100871, China}
\affiliation{Hefei National Laboratory, Hefei 230088, China}
\author{Yi Tan}
\affiliation{International Center for Quantum Materials and School of Physics, Peking University,
Beijing 100871, China}
\author{Wei Jia}
\affiliation{Key Laboratory of Quantum Theory and Applications of MoE, Lanzhou Center for Theoretical
Physics, and Key Laboratory of Theoretical Physics of Gansu Province, Lanzhou University,
Lanzhou 730000, China}
\author{Long Zhang}
\affiliation{School of Physics and Institute for Quantum Science and Engineering, Huazhong University
of Science and Technology, Wuhan 430074, China}
\affiliation{Hefei National Laboratory, Hefei 230088, China}
\author{Xiong-Jun Liu}
\thanks{Corresponding author: xiongjunliu@pku.edu.cn.}
\affiliation{International Center for Quantum Materials and School of Physics, Peking University,
Beijing 100871, China}
\affiliation{Hefei National Laboratory, Hefei 230088, China}
\affiliation{International Quantum Academy, Shenzhen 518048, China}
\affiliation{CAS Center for Excellence in Topological Quantum Computation, University of Chinese
Academy of Sciences, Beijing 100190, China}
\begin{abstract}
Three-dimensional 3rd-order topological insulators (TOTIs) and superconductors (TOTSCs),
as the highestorder topological phases hosting zero corner modes in physical dimension,
has sparked extensive research interest. However, such topological states have not
been discovered in reality due to the lack of experimental schemes of realization.
Here, we propose a novel surface Chern-Simons (CS) theory for 3rd-order topological
phases, and show that the theory enables a feasible and systematic design of TOTIs
and TOTSCs. We show that the emergence of zero Dirac (Majorana) corner modes is entirely
captured by an emergent $\mathbb{Z}_{2}$ CS term that can be further characterized
by a novel two-particle Wess-Zumino (WZ) term uncovered here in the surfaces of three-dimensional
topological materials. Importantly, our proposed CS term characterization and two-particle
WZ term mechanism provide a unique perspective to design TOTIs (TOTSCs) in terms
of minimal ingredients, feasibly guiding the search for underlying materials, with
promising candidates being discussed. This work shall advance both the theoretical
and experimental research for highest-order topological matters.
\end{abstract}
\maketitle
\textcolor{blue}{$\textit{Introduction}$.\textemdash }Higher-order topological phases,
an extension of conventional first-order topological phases \citep{HasanRMP2010,QiZhangRMP2011,BansilRMP2016},
have recently gained significant attention \citep{BenalcazarScience2017,BenalcazarPRB2017,LangbehnPRL2017,ZhidaSongPRL2017,SchindlerScience2018,GeierPRB2018,MiertPRB2018,EzawaPRB2018,CalugaruPRB2019,SchindlerJAP2020}.
Generally, the $n^{\mathrm{th}}$-order topological insulators (TIs) or superconductors
(TSCs) in $d$ dimension ($d$D) feature topologically protected gapless states on
its ($d-n$)D boundary, but are gapped elsewhere. In order to classify and identify
these exotic topological phases, various theoretical frameworks relying on detailed
bulk properties have been developed, such as nested Wilson loop \citep{BenalcazarScience2017,BenalcazarPRB2017},
symmetry indicators \citep{KhalafPRX2018,VishwanathNaturePhysics2019}, magnetic
topological quantum chemistry \citep{BernevigNatCommun2021} and other approaches
\citep{HughesPRB2019,JGongSciBull2021,LiPRB2021,XJLiuPRR2023}. The higher-order
topological phases are also intricately linked to broader research topics like non-Hermitian
physics \citep{KawabataPRB2020,OkugawaPRB2020,ZOUNC2021} and fracton orders \citep{YOUPRB2021}\textcolor{red}{. }

Till now, there has been a significant focus on 2nd-order cases \citep{SAYangPRL2019,ZFWangNanoLett2019,BJYangNPJ2020,NiuQianPRL2020,SAYangPRL2020,SAYangPRB2021,LiuFengPRResearch2021,ZFWangPRL2021,KhalafPRB2018,BernevigPRB2019,MatsugataniPRB2019,YXuPRB2019,XDaiPRL2019,PengPRL2019,QueirozPRL2019,ZYanPRL2019,SkurativskaPRR2020,KSunPRL2020,PlekhanovPRR2020,TakahashiPRR2020,RXZhangPRL2020,FZhangarXiv200514710,XLiuPRB2021,NMNguyenPRB2022}.
Specifically, the 2D 2nd-order TSCs have been extensively investigated \citep{LiuPRB2018,WangPRL2018,YanPRL2018,YanPRL2019,ZhuPRL2019,VolpezPRL2019,ZengPRL2019,PanPRL2019,ZhangPRL2019}
as Majorana corner modes exhibit non-Abelian statistics \citep{FWilczekNuclPhys1996,IvanovPhysRevLett2001,NayakPhysRevLett2005,MPFisherNatPhys2011,KTLawPhysRevX2014,PGaoPhysRevB2016,JSHongPhysRevB2022}
which is pivotal for topological quantum computation \citep{BomantaraPRB2020,PahomiPRR2020}.
Moreover, the experimental schemes have also been proposed to generate 3D 2nd-order
TIs (TSCs): applying Zeeman field \citep{CZhangPRL2020}, strain \citep{SchindlerScience2018}
and $s$-wave pairing \citep{SDSarmaPRL2019} to 3D parent topological phases can
induce gapless hinge modes. The materials, e.g. $\mathrm{bismuth}$ \citep{TNeupertNaturePhysics2018},
$\mathrm{WTe2}$ \citep{HChoiNatCommun2023} and $\mathrm{FeSe}$ \citep{KSBurchNanoLett2019}
have been identified experimentally to support 3D 2nd-order TIs and TSCs, respectively.
However, in contrast to 2nd-order topological phases, much less is known about 3rd-order
topological insulators (TOTIs) and superconductors (TOTSCs) \citep{BJYangPRB2019,BJYangPRR2020,ASahaPRB2021,VJuri=00010Di=000107PRB2021,HCPoPRX2021,JHBardarsonPRB2022,CNiunpjComputationalMater2022,CFangPhysRevB2022,XJLuoPhysRevB2023},
and even less is achieved in experiment. Currently, two critical issues confront
the thirdorder scenario and hinder the further development of this field. First,
while the TOTIs and TOTSCs can be predicted by the classification theories \citep{JGongSciBull2021,LiPRB2021,XJLiuPRR2023,BJYangPRB2019,CNiunpjComputationalMater2022},
their 3rd-order zero corner mode lacks an intuitive and unified characterization
theories that benefit the realization. Second, the few existing proposals \citep{BJYangPRR2020,ASahaPRB2021,VJuri=00010Di=000107PRB2021,JHBardarsonPRB2022,XJLuoPhysRevB2023}
for realizing TOTIs or TOTSCs are indeed short of physical feasibility and hard to
achieve in experiment, incapable of providing efficient guidance in searching for
underlying material candidates. 

In this letter, we address the above critical issues and develop a novel surface
Chern-Simons (CS) theory that is applicable generally to TOTIs and TOTSCs, and show
that the theory enables a feasible and intuitive principle in search for real material
candidates. We show that the presence of zero-energy corner modes in TOTIs and TOTSCs
is dictated by a $\mathbb{Z}_{2}$ CS term \citep{SRyuRevModPhys2016,CChanPhysRevLett2017},
which is defined within a synthetic 3D space and relies on two winding numbers derived
from surface mass fields and Dirac (or Majorana) cones. Further, we bring up an original
two-particle Wess-Zumino (WZ) term in the surfaces of 3D TIs (TSCs), which also characterizes
the Dirac and Majorana corner modes. The surface CS theory and two-particle WZ term
provide new and intuitive characterization of the 3rd-order zero corner modes and,
more importantly, inspire us to propose the realistic models of TOTI (TOTSC) based
on 2D surface construction. In particular, applying staggered (or uniform) in-plane
Zeeman fields and certain spin-orbit coupling (SOC) to conventional 3D TIs (TSCs)
drives the systems into 3rd-order phases with Dirac (Majorana) zero corner modes.
This facilitates to search for real materials, with promising candidates being identified. 

\begin{figure}[t]
\includegraphics[scale=0.93]{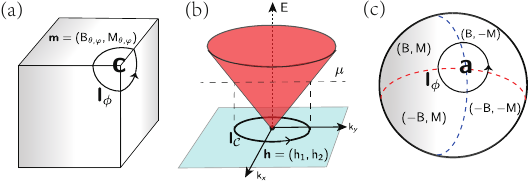} 

\caption{\label{fig:1} The CS characterization. (a) Illustration of the effective surface
mass field $\mathrm{\boldsymbol{m}}=(\mathrm{B}_{\theta,\varphi},\mathrm{M}_{\theta,\varphi})$.
Here, we employ a spherical coordinate frame with $\theta(\varphi)$ denoting the
polar (azimuth) angle. $\mathrm{\boldsymbol{l}}_{\phi}$ is a closed loop encircling
the target corner $\mathrm{\boldsymbol{c}}$. (b) Schematic of surface Dirac (Majorana)
cone governed by $\mathcal{H}=\mathrm{\boldsymbol{h}}(\boldsymbol{k})\cdot\mathrm{\boldsymbol{\Gamma}}$,
where the vector $\mathrm{\boldsymbol{h}}=(\mathrm{h}_{1},\mathrm{h}_{2})$ is a
linear function of $\boldsymbol{k}$ and $\mathrm{\boldsymbol{\Gamma}}$ represents
the gamma matrices. The loop $\mathrm{\boldsymbol{l}}_{\mathcal{C}}$ is obtained
by intersecting the Dirac (Majorana) cone with $\mu=(\mathrm{B}_{\phi}^{2}+\mathrm{M}_{\phi}^{2})^{1/2}$.
(c) Mass field with $\mathcal{W}_{\mathrm{\boldsymbol{m}}}=1$. We transform the
cubic sample in (a) into a sphere without gap closing, so corner $\mathrm{\boldsymbol{c}}$
is mapped to $\mathrm{\boldsymbol{a}}$, the intersection of two mass domain walls
(red and blue dashed lines). Remarkably, a zero-energy bound state at $\mathrm{\boldsymbol{a}}$
corresponds to a zero corner mode at $\mathrm{\boldsymbol{c}}$. }
\end{figure}

\textcolor{blue}{$\textit{ CS term characterization .}$$\text{\textemdash}$} We
start with the CS theory which enables a unified characterization of the zero corner
modes of TOTIs and TOTSCs. This theory is inspired by a previous conclusion \citep{CChanPhysRevLett2017}
that Majorana zero modes at the vortex cores in 2D superconductors are generically
protected by an emerging CS invariant $(-1)^{\nu_{3}}=\sum_{i}n_{i}w_{i}$, where
$w_{i}$ denotes the phase winding of superconducting order projected onto $i$-th
FS and $n_{i}$ is the integer vortex winding number (vorticity). A single Majorana
zero mode is protected when the index $\nu_{3}=1$. For TOTIs and TOTSCs, it's important
to highlight that the 3rd-order zero corner modes, which can be viewed as a unique
type of defect-bound states, share similarities with vortex core modes. We then expect
that the zero corner states may also be characterized by a similar CS term $\tilde{\theta}$
resembling the framework of 2D theory. 

Thereinafter, we derive the CS term $\tilde{\theta}$ for the zero corner modes through
an effective surface theory. The surface effective Hamiltonian of the TOTIs (TOTSCs)
can be described generally as $\mathcal{H}_{\mathrm{surf}}^{3\mathrm{rd}}=\mathrm{h}_{1}(\boldsymbol{k})\Gamma_{1}+\mathrm{h}_{2}(\boldsymbol{k})\Gamma_{2}+\mathrm{B}_{\theta,\varphi}\Gamma_{3}+\mathrm{M}_{\theta,\varphi}\Gamma_{4}$
with $\{\Gamma_{i},\Gamma_{j}\}=2\delta_{ij}$, where $\mathcal{H}=\mathrm{h}_{1}(\boldsymbol{k})\Gamma_{1}+\mathrm{h}_{2}(\boldsymbol{k})\Gamma_{2}$
represents the effective Hamiltonian of surface Dirac (Majorana) cones and $\mathrm{\boldsymbol{m}}=(\mathrm{B}_{\theta,\varphi},\mathrm{M}_{\theta,\varphi})$
constitutes the surface mass field {[}Fig.\ref{fig:1}(a){]}. It is known that $\tilde{\theta}$
can be defined on a manifold $k^{d}\times\mathcal{M}^{D}$, with $d+D$ being odd
\citep{SRyuRevModPhys2016}. To address this requirement, apart from 2D surface $\boldsymbol{k}$-space,
we also focus on the loop $\mathrm{\boldsymbol{l}}_{\phi}$ {[}solid line in Fig.\ref{fig:1}(a){]}
surrounding target corner $\mathrm{\boldsymbol{c}}$ and take real space parameter
$\phi$ as a synthetic dimension of ring geometry $\mathcal{S}^{1}$. Then the surface
Hamiltonian can be expressed in a compact 3D space $k^{3}=k^{2}\times\mathcal{S}^{1}$
spanned by $(\boldsymbol{k},\phi)$ as follows: $\mathcal{H}_{\mathrm{surf}}^{3\mathrm{rd}}=\mathrm{h}_{1}(\boldsymbol{k})\Gamma_{1}+\mathrm{h}_{2}(\boldsymbol{k})\Gamma_{2}+\mathrm{B}_{\phi}\Gamma_{3}+\mathrm{M}_{\phi}\Gamma_{4}$.
And the Chern-Simons term is represented as $\tilde{\theta}=\frac{1}{4\pi}\int\mathrm{d}^{3}k\epsilon_{\mathrm{abc}}\mathrm{tr}[\mathcal{A}_{\mathrm{a}}\partial_{\mathrm{b}}\mathcal{A}_{\mathrm{c}}+\mathrm{i}\frac{2}{3}\mathcal{A}_{\mathrm{a}}\mathcal{A}_{\mathrm{b}}\mathcal{A}_{\mathrm{c}}]$
where the non-Abelian Berry connection reads $\mathcal{A}_{\mathrm{a}=(\boldsymbol{k},\phi)}=-\mathrm{i}\tilde{\langle n|}\partial_{a}\tilde{|n^{\prime}\rangle}$
and $|\tilde{n}\rangle$ represents the degenerate eigenvector of occupied bands
$-E=-(\mathrm{h}_{1}^{2}+\mathrm{h}_{2}^{2}+\mathrm{B}_{\phi}^{2}+\mathrm{M}_{\phi}^{2})^{1/2}$.
We show that \citep{SM}
\begin{equation}
\tilde{\theta}=\pi\mathcal{W}_{\mathrm{\boldsymbol{h}}}\mathcal{W}_{\mathrm{\boldsymbol{m}}},\label{eq:1}
\end{equation}
where $\mathcal{W}_{\mathrm{\boldsymbol{h}}}=\frac{1}{2\pi}\oint_{\mathrm{\boldsymbol{l}}_{\mathcal{C}}}\mathrm{d}\mathrm{\boldsymbol{l}}\cdot(\mathrm{h}_{2}\nabla_{\mathrm{\boldsymbol{k}}}\mathrm{h}_{1}-\mathrm{h}_{1}\nabla_{\mathrm{\boldsymbol{k}}}\mathrm{h}_{2})/(\mathrm{h}_{2}^{2}+\mathrm{h}_{1}^{2})$
and $\mathcal{W}_{\mathrm{\boldsymbol{m}}}=\frac{1}{2\pi}\oint_{\mathrm{\boldsymbol{l}}_{\phi}}\mathrm{d}\phi(\mathrm{M}_{\phi}\partial_{\phi}\mathrm{B}_{\phi}-\mathrm{B}_{\phi}\partial_{\phi}\mathrm{M}_{\phi})/(\mathrm{B}_{\phi}^{2}+\mathrm{M}_{\phi}^{2})$
represent the winding number for vector $\boldsymbol{\mathrm{h}}=(\mathrm{h}_{1},\mathrm{h}_{2})$
along contour $\mathrm{\boldsymbol{l}}_{\mathcal{C}}$ {[}Fig.\ref{fig:1}(b){]}
and surface mass field $\mathrm{\boldsymbol{m}}$ on loop $\mathrm{\boldsymbol{l}}_{\phi}$,
respectively. If the surface Dirac (Majorana) cones dictated by $\mathcal{H}=\mathrm{h}_{1}\Gamma_{1}+\mathrm{h}_{2}\Gamma_{2}$
contribute to $\mathcal{W}_{\mathrm{\boldsymbol{h}}}=1$, as in usual TIs and TSCs,
the CS term becomes $\tilde{\theta}=\pi\mathcal{W}_{\mathrm{\boldsymbol{m}}}$, indicating
that $\tilde{\theta}$ can be solely determined by the surface mass field. 

We show that a nonzero $\tilde{\theta}$ corresponds to a zero corner mode (see details
in Supplementary Materials \citep{SM}). Fig.\ref{fig:1}(c) illustrates a nontrivial
surface mass field, which exhibits a nonzero winding number $\mathcal{W}_{\mathrm{\boldsymbol{m}}}=1$,
and then $\tilde{\theta}=\pi$ . In this case $\mathcal{H}_{\mathrm{surf}}^{3\mathrm{rd}}$
resembles a 2D Dirac Hamiltonian with a vortex profile which binds a zero-energy
bound state at the core, i.e. the target corner $\mathrm{\boldsymbol{c}}$, indicating
that it belongs to the symmetry class D \citep{KhalafPRB2018,SRyuRevModPhys2016,CChanPhysRevLett2017}.
Instead, no corner mode exists if the mass winding $\mathcal{W}_{\mathrm{\boldsymbol{m}}}=0$
\citep{SM}. Therefore, a nonzero $\tilde{\theta}$ $(\pm\pi)$ signifies the presence
of zero corner mode of TOTIs or TOTSCs. 

\textcolor{blue}{$\textit{ Two-particle WZ term.}$$\text{\textemdash}$} A more
intuitive characterization of the zero corner modes can be obtained by introducing
an original topological invariant dubbed \textit{Two-particle WZ term}, which is
defined in the surfaces of the parent 3D TIs (TSCs). We show that the presence of
zero corner modes can be contingent upon this term under specific conditions. 

\begin{figure}[t]
\includegraphics[scale=0.9]{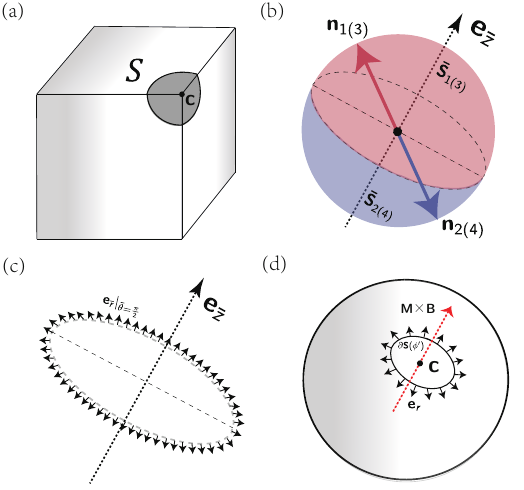}

\caption{\label{fig:2} The two-particle WZ mechanism. (a) Area $\mathcal{\boldsymbol{S}}$
centered at the target corner $\mathrm{\boldsymbol{\mathrm{\boldsymbol{c}}}}$. (b)
(Pseudo)spin textures within $\mathcal{\boldsymbol{S}}$ in (pseudo)spin space. Due
to the ``double-TI(TSC) construction'', we define two pairs of (pseudo)spin polarizations
$\{\mathrm{\boldsymbol{n}}_{1},\mathrm{\boldsymbol{n}}_{2}\}$ and $\{\mathrm{\boldsymbol{n}}_{3},\mathrm{\boldsymbol{n}}_{4}\}$
with $\mathrm{\boldsymbol{n}}_{1}=-\mathrm{\boldsymbol{n}}_{2}=\mathrm{\boldsymbol{n}}_{3}=-\mathrm{\boldsymbol{n}}_{4}=\boldsymbol{e}_{r}$.
Ideally, each group of (pseudo)spins form perfect unit spherical surfaces $\mathcal{\boldsymbol{\bar{S}}}_{1(3)}+\mathcal{\bar{\boldsymbol{S}}}_{2(4)}$
in (pseudo)spin space, and the total (pseudo)spin textures coincide. Note that the
vector $\boldsymbol{e}_{r=r_{\mathrm{\boldsymbol{c}}}}$ is required to be the $\boldsymbol{e}_{\bar{z}}$
in (pseudo)spin space. (c) The winding of $\boldsymbol{\boldsymbol{e}_{\bar{r}}}|_{\bar{\theta}=\frac{\pi}{2}}$
along the equator of $\mathcal{\boldsymbol{\bar{S}}}_{1(3)}+\mathcal{\bar{\boldsymbol{S}}}_{2(4)}$.
(d) The winding of $\boldsymbol{e}_{r}$ on the boundary $\partial\mathcal{\boldsymbol{S}}$
encircling the corner $\mathrm{\boldsymbol{c}}$. The vector $\mathrm{\boldsymbol{M}}\times\mathrm{\boldsymbol{B}}$
is parallel to $\boldsymbol{e}_{r=r_{\mathrm{\boldsymbol{c}}}}$.}
\end{figure}

According to the ``double-TI(TSC) construction'' \citep{KhalafPRX2018,KhalafPRB2018,BernevigPRB2019},
we first stack two identical 3D time-reversal invariant TIs (TSCs) to construct a
3D total TIs (TSCs): $\mathcal{H}_{\mathrm{TI(TSC),total}}(\boldsymbol{k})=\mathcal{H}_{\mathrm{TI(TSC),\alpha}}(\boldsymbol{k})\oplus\mathcal{H}_{\mathrm{TI(TSC),\beta}}(\boldsymbol{k})$,
which yields two pairs of topological surface states $\{|\varepsilon_{(1,2)}(\boldsymbol{r})\rangle\}_{\alpha};\{|\varepsilon_{(3,4)}(\boldsymbol{r})\rangle\}_{\beta}$.
Here we can define the surface (pseudo)spin polarizations: $\mathrm{\boldsymbol{n}}_{\nu=(1,2,3,4)}(\boldsymbol{r})=\langle\varepsilon_{\nu}(\boldsymbol{r})|\boldsymbol{s}|\varepsilon_{\nu}(\boldsymbol{r})\rangle$
\citep{XJLiuPRL2014,CXLiuPRB2016,XJLiuPRB2022}, where Pauli matrix $\boldsymbol{s}$
represents the spin degrees of freedom. Generally, (pseudo)spins are taken as the
surface normal vector $\boldsymbol{e}_{r}$ of 3D time-reversal invariant TIs (TSCs).
We focus on an area $\mathcal{\boldsymbol{S}}$ of the surface of total TIs (TSCs),
centered at corner $\mathrm{\boldsymbol{c}}$ {[}Fig.\ref{fig:2}(a){]}, and study
the corresponding surface (pseudo)spin textures defined on it. As shown in Fig.\ref{fig:2}(b),
a single $\mathrm{\boldsymbol{n}}_{\boldsymbol{\nu}=(1,2,3,4)}$ within $\mathcal{\boldsymbol{S}}$
cannot form a complete sphere, but the joint textures of two polarizations $\mathrm{\boldsymbol{n}}_{\boldsymbol{1}}$
and $\mathrm{\boldsymbol{n}}_{\boldsymbol{2}}$ ($\boldsymbol{1}=1,3;\boldsymbol{2}=2,4$)
can constitute a ``two-particle'' unit sphere $\boldsymbol{\bar{S}}_{1}+\boldsymbol{\bar{S}}_{2}$
($\boldsymbol{\bar{S}}_{3}+\boldsymbol{\bar{S}}_{4}$) in (pseudo)spin space under
ideal condition, resembling a hedgehog ball. Thus we can define a novel \textit{two-particle
WZ term} on $\mathcal{\boldsymbol{S}}$: 
\begin{equation}
\mathrm{\tilde{n}}_{\mathrm{WZ}}^{\alpha}=\mathrm{\tilde{n}}_{\mathrm{WZ}}^{\beta}=\int_{\mathcal{\boldsymbol{S}}}\frac{\mathrm{d}\theta\mathrm{d}\varphi}{2\pi}(\partial_{\theta}\tilde{\mathcal{A}}_{\varphi}-\partial_{\varphi}\tilde{\mathcal{A}}_{\theta}),\label{eq:2}
\end{equation}
where $\tilde{\mathcal{A}}_{\theta(\varphi)}=-\mathrm{i}\langle\varepsilon_{\mathrm{\boldsymbol{1}}}|\otimes\langle\varepsilon_{\mathrm{\boldsymbol{2}}}|\partial_{\theta(\varphi)}|\varepsilon_{\mathrm{\boldsymbol{1}}}\rangle\otimes|\varepsilon_{\boldsymbol{2}}\rangle$
represents two-particle Berry connection, with $\theta(\varphi)$ being polar (azimuthal)
angle in real space. As displayed in Fig.\ref{fig:2}(c), we find that Eq.(\ref{eq:2})
can further reduce to a winding number of the normal vector $\boldsymbol{\boldsymbol{e}_{\bar{r}}}$
along the equator of unit sphere $\boldsymbol{\bar{S}}_{1}+\boldsymbol{\bar{S}}_{2}(\boldsymbol{\bar{S}}_{3}+\boldsymbol{\bar{S}}_{4})$:
$\mathrm{\tilde{n}}_{\mathrm{WZ}}^{\alpha}=\mathrm{\tilde{n}}_{\mathrm{WZ}}^{\beta}=\oint_{\mathrm{equator}}\frac{\mathrm{d}\bar{\varphi}}{2\pi}[\boldsymbol{e_{\bar{z}}}\cdot(\boldsymbol{\boldsymbol{e}_{\bar{r}}}|_{\bar{\theta}=\frac{\pi}{2}}\times\partial_{\bar{\varphi}}\boldsymbol{\boldsymbol{e}_{\bar{r}}}|_{\bar{\theta}=\frac{\pi}{2}})]$,
where $\bar{\theta}(\bar{\varphi})$ denotes the polar (azimuthal) angle in (pseudo)spin
space with respect to base vector $\boldsymbol{e}_{\bar{z}}$ which is chosen as
the vector $\boldsymbol{e}_{r=r_{\mathrm{\boldsymbol{c}}}}$ from ``center of $\partial\mathcal{\boldsymbol{S}}$\textquotedblright{}
\citep{SM} to corner $\mathrm{\boldsymbol{c}}$. Notice that $\mathrm{\tilde{n}}_{\mathrm{WZ}}$
is an intuitive topological invariant that emerges from (pseudo)spin textures around
target corner and contains the necessary surface information used for identifying
the corner modes of TOTIs and TOTSCs. 

We then apply a certain surface mass field $\mathrm{\boldsymbol{m}}=(\boldsymbol{e}_{r}\cdot\boldsymbol{\mathrm{B}},\boldsymbol{e}_{r}\cdot\boldsymbol{\mathrm{M}})$
($\boldsymbol{\mathrm{B}}$ and $\boldsymbol{\mathrm{M}}$ are constant unit vectors
with $\boldsymbol{\mathrm{M}}\times\boldsymbol{\mathrm{B}}\neq0$) to gap out the
surface of total TIs (TSCs) to achieve the TOTIs (TOTSCs). The CS term about corner
$\mathrm{\boldsymbol{c}}$ reads $\tilde{\theta}=\oint_{\mathcal{\partial\boldsymbol{S}}(\phi^{\prime})}\frac{\mathrm{d}\phi^{\prime}}{2}|\boldsymbol{\mathrm{M}}\times\boldsymbol{\mathrm{B}}|[\boldsymbol{e}_{\boldsymbol{\mathrm{M}}\times\boldsymbol{\mathrm{B}}}\cdot(\boldsymbol{\boldsymbol{e}_{r}}\times\partial_{\phi^{\prime}}\boldsymbol{\boldsymbol{e}_{r}})]/[(\boldsymbol{e}_{r}\cdot\boldsymbol{\mathrm{B}})^{2}+(\boldsymbol{e}_{r}\cdot\boldsymbol{\mathrm{B}})^{2}]$,
which is also related to the winding of $\boldsymbol{\boldsymbol{e}_{r}}$ {[}Fig.\ref{fig:2}(d){]}. 

Combining the $\mathrm{\tilde{n}}_{\mathrm{WZ}}$ and $\tilde{\theta}$ derived above,
we reach that \citep{SM}
\begin{equation}
|\tilde{\theta}|=\pi|\mathrm{\tilde{n}}_{\mathrm{WZ}}^{\alpha}|=\pi|\mathrm{\tilde{n}}_{\mathrm{WZ}}^{\beta}|,\label{eq:3}
\end{equation}
which is a key result that the two-particle WZ term leads to a non-zero CS term in
the presence of the mass field $(\boldsymbol{e}_{r}\cdot\boldsymbol{\mathrm{B}},\boldsymbol{e}_{r}\cdot\boldsymbol{\mathrm{M}})$.
Thus, if the surface (pseudo)spin textures around the target corner has a nonzero
two-particle WZ term $(\mathrm{\tilde{n}}_{\mathrm{WZ}}^{\alpha}=\mathrm{\tilde{n}}_{\mathrm{WZ}}^{\beta}=1)$,
a topological zero corner mode emerges whenever the mass field $(\boldsymbol{e}_{r}\cdot\boldsymbol{\mathrm{B}},\boldsymbol{e}_{r}\cdot\boldsymbol{\mathrm{M}})$
is introduced. We note that the (pseudo)spin textures within $\mathcal{S}$ may not
constitute a perfect unit spherical surface, but can be connected to a unit spherical
surface through the \textquotedbl continuation\textquotedbl{} which does not bring
about additional nodes of mass. A non-zero $\mathrm{\tilde{n}}_{\mathrm{WZ}}$ is
obtained as long as the actual surface (pseudo)spin textures cover the area where
the mass nodes are obtained \citep{SM}. 

Before proceeding, we clarify two points about the surface Chern-Simons theory. ($\mathrm{I}$)
The characterization theory based on $\tilde{\theta}$ is built on the general effective
surface Hamiltonian rather than detailed bulk properties \citep{JGongSciBull2021,LiPRB2021,XJLiuPRR2023,BJYangPRB2019,CNiunpjComputationalMater2022},
making it applicable to generic TOTIs and TOTSCs, irrespective of the concrete methods
for achieving these phases. ($\mathrm{II}$) The $\mathrm{\tilde{n}}_{\mathrm{WZ}}$
mechanism shows a rather intuitive physical characterization: the corner modes are
associated with the surface (pseudo)spin textures that are governed by the mass field.
These features provide a powerful guidance in constructing the physical models and
facilitate the search for real materials. 

\begin{figure}[t]
\includegraphics[scale=0.9]{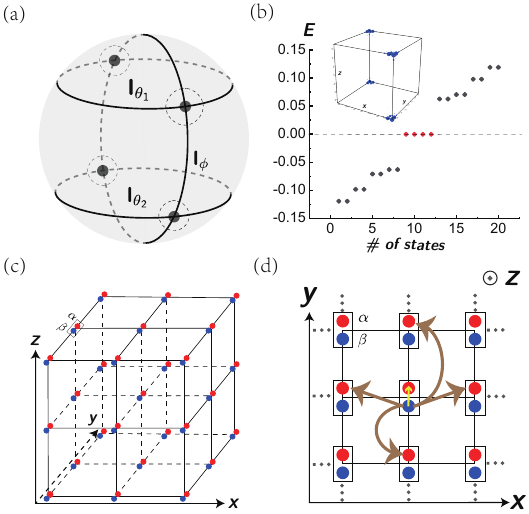}

\caption{\label{fig:3} The TOTSC model is described by $\tilde{\theta}$ characterization
with parameters $\mu=2$, $m_{\ast}=0.5$, $\Delta_{p}/k_{F}=1$, $\lambda_{so,0}=1$,
$\lambda_{so,1}=0.8$, and $|\mathrm{\boldsymbol{B}}_{\parallel}|=1$. (a) The designed
square lattice for this physical TOTSC model features $\alpha$ and $\beta$ orbitals
within a unit cell, represented by red and blue dots, respectively. (b) The schematic
illustration of $\mathcal{T}$-invariant spin-orbit coupling in real space, comprising
on-site (yellow arrow) and nearest-neighbouring (brown arrow) terms. (c) Predicted
distribution of surface effective mass field vortices by $\tilde{\theta}$ characterization.
$\mathrm{\boldsymbol{l}}_{\phi}$ and $\mathrm{\boldsymbol{l}}_{\theta_{1(2)}}$
represent mass domain walls resulting from the Zeeman and SOC terms, respectively.
(d) Numerical results concerning TOTSCs, showing consistent numbers and positions
of Majorana zero corner modes with theoretical predictions from Chern-Simons term
$\tilde{\theta}$ characterization. }
\end{figure}

\begin{figure}[t]
\includegraphics[scale=0.9]{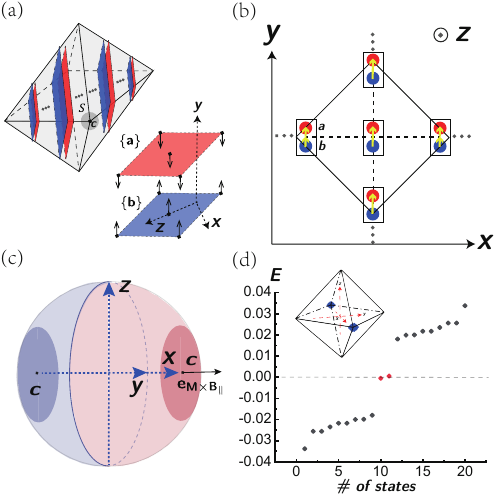}

\caption{\label{fig:4} The $\tilde{\mathrm{n}}_{\mathrm{WZ}}$ mechanism is applied to the
physical model of TOTIs with parameters $m=2$, $t=1$, $\lambda_{so}=1$, and $\mathrm{B}=1$.
(a) Sample geometry illustration: $a$ and $b$ represent different sublattice sites
within a unit cell, and the corresponding layers $\{a\}$ and $\{b\}$ are alternately
arranged. Antiferromagnetism can be induced between $\{a\}$ and $\{b\}$. (b) Visualization
of an alternative time-reversal invariant SOC in real space, featuring only the intra-cell
term (yellow arrow) remaining. (c) (Pseudo)spin textures (deep red and deep blue
regions) around the target corner $\boldsymbol{c}$. Although perfect spherical surfaces
cannot be obtained from these textures, the $\tilde{\mathrm{n}}_{\mathrm{WZ}}$ is
still well-defined through \textquotedbl continuation\textquotedbl{} (light red
and light blue regions denote auxiliary textures) due to $\boldsymbol{e}_{\mathrm{\boldsymbol{M}\times\boldsymbol{B}_{\parallel}}}=\boldsymbol{e}_{r=r_{\boldsymbol{c}}}$.
(d) Numerical results validate the theoretical predictions.}
\end{figure}

${\color{red}{\color{blue}\textit{Physical realizations}.\text{\textemdash}}}$ The
surface Chern-Simons theory offers a unique perspective for the emergence of zero
corner modes\textemdash it opens new pathways to develop realistic physical models
for TOTIs (TOTSCs) via the mass field construction. In the following, we shall introduce
implementation strategies for TOTSCs and TOTIs in terms of $\tilde{\theta}$ characterization
and $\mathrm{\tilde{n}}_{\mathrm{WZ}}$ mechanism, respectively, with which the promising
candidates will be discussed. Now let us begin with 
\begin{equation}
\mathcal{H}_{\mathrm{TOTSC}}(\boldsymbol{k})=\mathcal{H}_{\mathrm{TSC}}^{\mathrm{total}}(\boldsymbol{k})+\mathrm{\boldsymbol{B}_{\parallel}}\cdot\boldsymbol{s}+\lambda_{so}(\boldsymbol{k})\sigma_{y}s_{z},\label{eq:4}
\end{equation}
where $\lambda_{so}(\boldsymbol{k})=\lambda_{so,0}-\lambda_{so,1}(\cos k_{x}+\cos k_{y})$.
The total TSC comprises two copies of TSCs $(\mathcal{H}_{\mathrm{TSC,\alpha}}\oplus\mathcal{H}_{\mathrm{TSC,\beta}})$
with the same winding numbers \citep{SRyuRevModPhys2016} and the Hamiltonian is
given by $\mathcal{H}_{\mathrm{TSC},\mathrm{total}}(\boldsymbol{k})=(\Delta_{p}/k_{F})\sum_{j=x,y}\sin k_{j}\sigma_{0}\tau_{x}s_{j}+(\Delta_{p}/k_{F})\sin k_{z}\sigma_{z}\tau_{x}s_{z}-[\mu-3/m_{\ast}+(1/m_{\ast})\sum_{j=x,y,z}\cos k_{j}]\sigma_{z}\tau_{z}s_{0}$.
Here, $\boldsymbol{\sigma}$, $\boldsymbol{\tau}$ and $\boldsymbol{s}$ denote the
orbital, Nambu and spin degrees of freedom, respectively. As shown in Fig.\ref{fig:3}(a),
two orbitals $(\alpha,\beta$) are depicted at each site. The mass field is given
by the last two terms of Eq.(\ref{fig:4}), namely, the Zeeman term originated from
an in-plane magnetic field $\mathrm{\boldsymbol{B}_{\parallel}}=(1/\sqrt{2})(\boldsymbol{e}_{x}+\boldsymbol{e}_{y})$
and the SOC term preserving time-reversal symmetry: $\hat{\mathcal{T}}^{-1}\lambda_{so}(\boldsymbol{k})\sigma_{y}s_{z}\hat{\mathcal{T}}=\lambda_{so}(-\boldsymbol{k})\sigma_{y}s_{z}$.
Note that the SOC term, involving the real-space on-site and nearest-neighbour parts
shown in Fig.\ref{fig:3}(b), is captured by $\hat{H}_{\mathrm{SOC}}^{\mathrm{I}}=\sum_{\boldsymbol{i}}-\mathrm{i}\lambda_{so,0}(\hat{c}_{\boldsymbol{i},\alpha}^{\dagger}s_{z}\hat{c}_{\boldsymbol{i},\beta}-\hat{c}_{\boldsymbol{i},\beta}^{\dagger}s_{z}\hat{c}_{\boldsymbol{i},\alpha})+\sum_{\langle\boldsymbol{i},\boldsymbol{j}\rangle}(\frac{\mathrm{i}\lambda_{so,1}}{2}\hat{c}_{\boldsymbol{i},\alpha}^{\dagger}s_{z}\hat{c}_{\boldsymbol{j},\beta}+\mathrm{h.c.})$
\citep{FLiuNatCommun2016,HHuangPhysRevLett2022}, where $\lambda_{so,0}$ and $\lambda_{so,1}$
represent the corresponding SOC strengths. 

Using the $\tilde{\theta}$ characterization, we can predict Majorana zero corner
modes. The surface effective mass field is given by $\mathrm{\boldsymbol{m}}=(\mathrm{m}_{\phi},\mathrm{m}_{\theta})$
\citep{SM}. Here $\mathrm{m}_{\phi}$ is induced by the magnetic field that is confined
to the $x-y$ plane, and the mass domain wall of $\mathrm{m}_{\phi}$ spreads along
one meridian $\mathrm{\boldsymbol{l}}_{\phi}$. Meanwhile, $\mathrm{m}_{\theta}$
induced by $\mathcal{H}_{\mathrm{SOC}}(\boldsymbol{k})$ creates two additional domain
walls at parallels $\mathrm{\boldsymbol{l}}_{\theta_{1}}$ and $\mathrm{\boldsymbol{l}}_{\theta_{2}}$.
The intersections of $\mathrm{\boldsymbol{l}}_{\phi}$ and $\mathrm{\boldsymbol{l}}_{\theta_{1(2)}}$
in Fig.\ref{fig:3}(c) represent mass field vortex cores with $\mathcal{W}_{\mathrm{\boldsymbol{m}}}=1$.
Thus the TOTSC hosts four Majorana zero corner modes. Numerical results shown in
Fig.\ref{fig:3}(d) support this prediction.  

The proposal for TOTIs follows a similar approach. We consider the Hamiltonian 
\begin{align}
\mathcal{H}_{\mathrm{TOTI}}(\boldsymbol{k})= & \mathcal{H}_{\mathrm{TI}}^{\mathrm{total}}(\boldsymbol{k})+\rho_{z}\mathrm{\boldsymbol{B}_{\parallel}}\cdot\boldsymbol{s}+\rho_{y}\mathrm{\boldsymbol{M}}\cdot\boldsymbol{s},\label{eq:5}
\end{align}
where the total TI Hamiltonian reads $\mathcal{H}_{\mathrm{TI}}^{\mathrm{total}}(\boldsymbol{k})=\mathcal{H}_{\mathrm{TI}}^{a}(\boldsymbol{k})\oplus\mathcal{H}_{\mathrm{TI}}^{b}(\boldsymbol{k})=(m-t\sum_{i}\cos k_{i})\rho_{0}s_{0}\sigma_{z}+t(\sum_{i}\sin k_{i}\rho_{0}s_{i})\sigma_{x}$,
and $\boldsymbol{\rho}$ denotes the Pauli matrix for sublattice degrees of freedom.
We explore the constant terms ($\rho_{z}\mathrm{\boldsymbol{B}_{\parallel}}\cdot\boldsymbol{s}$,
$\rho_{y}\mathrm{\boldsymbol{M}}\cdot\boldsymbol{s}$) with $\mathrm{\boldsymbol{B}_{\parallel}}=-\mathrm{B}\boldsymbol{e}_{y}$
and $\boldsymbol{\mathrm{M}}=\lambda_{so}\boldsymbol{e}_{z}$. The staggered Zeeman
term $\rho_{z}\mathrm{\boldsymbol{B}_{\parallel}}\cdot\boldsymbol{s}$ can be implemented
by applying antiferromagnetism between layers $\{a\}$ and $\{b\}$ as plotted in
Fig.\ref{fig:4}(a), while the $\mathcal{\hat{T}}$-invariant SOC shown in Fig.\ref{fig:4}(b)
only remains the intra-unit cell term in the real space: $\hat{H}_{\mathrm{SOC}}^{\mathrm{II}}=\sum_{\boldsymbol{i}}-\mathrm{i}\lambda_{so}(\hat{c}_{\boldsymbol{i},a}^{\dagger}s_{z}\hat{c}_{\boldsymbol{i},b}-\hat{c}_{\boldsymbol{i},b}^{\dagger}s_{z}\hat{c}_{\boldsymbol{i},a})$.
Here $a$ and $b$ denote different sublattice sites within a unit cell. 

Remarkably, the two constant terms ($\rho_{z}\mathrm{\boldsymbol{B}_{\parallel}}\cdot\boldsymbol{s},\rho_{y}\mathrm{\boldsymbol{M}}\cdot\boldsymbol{s}$)
generate the desired surface mass field ($\boldsymbol{e}_{r}\cdot\mathrm{\boldsymbol{B}_{\parallel}}$,$\boldsymbol{e}_{r}\cdot\boldsymbol{\mathrm{M}}$),
and give rise to the corner modes of the TOTI according to the $\tilde{\mathrm{n}}_{\mathrm{WZ}}$
mechanism. As illustrated in Fig.\ref{fig:4}(c), the (pseudo)spin textures around
$\mathrm{\boldsymbol{c}}$ are actually not closed. However, we can find that the
well-defined two-particle WZ term ($\tilde{\mathrm{n}}_{\mathrm{WZ}}^{a}=\tilde{\mathrm{n}}_{\mathrm{WZ}}^{b}=1$)
is obtained by \textquotedblleft continuation\textquotedblright{} as $\boldsymbol{e}_{\boldsymbol{\mathrm{M}}\times\mathrm{\boldsymbol{B}_{\parallel}}}\parallel\boldsymbol{e}_{x}$
which does not involve mass nodes \citep{SM}. In consequence, a zero corner mode
is obtained at the corner $\mathrm{\boldsymbol{c}}$. The numerical results in Fig.\ref{fig:4}(d)
confirm this analysis. 

With the above physical schemes we can now propose the promising candidates for realization.
In order to achieve the TOTSCs described by Eq.(\ref{eq:4}), besides the applied
Zeeman field in the $x-y$ plane, there should be at least two different orbitals
$(\alpha,\beta)$ for each lattice site, with which the $\hat{H}_{\mathrm{SOC}}^{\mathrm{I}}$
can be implemented. For our purpose the typical orbitals $(\alpha,\beta)$ can be
($p_{x},p_{y}$), ($d_{xz},d_{yz}$), ($d_{xy},d_{x^{2}-y^{2}}$). We then find a
promising candidate based on the superconducting material $\beta$-$\mathrm{PdBi_{2}}$
\citep{MMSharmaSuperconductorScienceandTechnology} which exhibits $p$-wave pairing
\citep{YFLiScience2019,AKolapoScientificReports2019} and its topological surface
states \citep{MSakanoNatureCommunications2015} meets the condition as the Fermi
surface is mostly contributed from $\mathrm{Bi}$ $(6p_{x},6p_{y})$ orbitals \citep{IRSheinJournalSuperconductivityNovelMagnetism2013}.
Thus the TOTSC could be realized by applying Zeeman field. As for the TOTIs in Eq.(\ref{eq:5}),
there are also two key ingredients, the staggered Zeeman term $\rho_{z}\mathrm{\boldsymbol{B}_{\parallel}}\cdot\boldsymbol{s}$
and $\hat{H}_{\mathrm{SOC}}^{\mathrm{II}}$, in reaching this topological phase.
Notably, the 3D antiferromagnetic (AFM) TI \citep{MoorePRB2010} can be an ideal
platform to implement these terms since the staggered Zeeman field are naturally
captured by the interlayer antiferromagnetism orders and the $\hat{H}_{\mathrm{SOC}}^{\mathrm{II}}$
is very promising to be realized by inducing the spin-dependent couplings between
neighbouring AFM layers. With this we conjecture that in the 3D AFM topological materials
(e.g. $\mathrm{MnBi_{2}Te_{4}}$ \citep{JWangPhysRevLett2019,HBentmannNature2019,XieNSR})
introducing the spin-dependent interlayer couplings should generate the desired TOTI
phases \citep{SM}. We note that while to further confirm the proposed candidates
with more concrete details necessitates more sophisticated numerical methods \citep{XWanNature2019,BABernevigNature2020},
our present theory has provided clear guidance in search for the 3rd-order TIs and
TSCs in real materials.

\textcolor{blue}{$\textit{Conclusion}.$\textemdash{}}\textcolor{red}{{} }We have proposed
the surface Chern-Simons theory for TOTIs and TOTSCs, which provides a unified and
intuitive characterization of the Dirac and Majorana corner modes, and facilitates
the physical realizations of these 3rdorder zero modes. We showed that the CS term,
dictating the existence of zero corner modes in all TOTIs and TOTSCs, is connected
to mass field and Dirac (Majorana) cone winding numbers, and further to a novel two-particle
WZ term derived from surface (pseudo)spin textures of the parent 3D topological materials.
This intuitive topological characterization offers a fresh approach for realizing
physical models of TOTIs (TOTSCs) and aid the search for real materials, of which
the promising candidates have been discussed. This study opens up a clear route to
discover 3rd-order topological matter in both theory and experiment. 

$\textit{Acknowledgment}s$.\textemdash{} We thank Xin-Chi Zhou, Ting-Fung Jeffrey
Poon, and Bao-Zong Wang for valuable discussions. This work was supported by National
Key Research and Development Program of China (No. 2021YFA1400900), the National
Natural Science Foundation of China (Grants No. 12261160368, No. 11825401, No. 12104205,
and No. 11921005), and the Innovation Program for Quantum Science and Technology
(Grant No. 2021ZD0302000).

%%%%%%%%%%%%%%%%%%%%%%%%%%%%%%%%%%%%%% %%   Supplementary Material %%%%%%%%%%%%%%%%%%%%%%%%%%%%%%%%%%%%%%
\renewcommand{\thesection}{S-\arabic{section}} 
\setcounter{section}{0}  %  this will re-count section from 1 
\renewcommand{\theequation}{S\arabic{equation}} 
\setcounter{equation}{0}  %  this will re-count eq from 1 
\renewcommand{\thefigure}{S\arabic{figure}} 
\setcounter{figure}{0}  %  this will re-count eq from 1 
\renewcommand{\thetable}{S\Roman{table}} 
\setcounter{table}{0}  %  this will re-count eq from 1
\onecolumngrid \flushbottom %\onecolumn

\newpage
\begin{center}\large \textbf{Supplementary Material} \end{center}

In this supplementary material, we first compute the Chern-Simons (CS) term $\tilde{\theta}$
analytically and demonstrate that a nonzero $\tilde{\theta}$ can correspond to a
zero corner mode (\ref{sec:Chern-Simons-term-}). Next we derive the two-particle
Wess-Zumino (WZ) term $\tilde{\mathrm{n}}_{\mathrm{WZ}}$ based on (pseudo)spin textures
defined on the surface of 3D topological insulators (TIs) or superconductors (TSCs)
and introduce the $\tilde{\mathrm{n}}_{\mathrm{WZ}}$ mechanism for closed and non-closed
(pseudo)spin textures case in detail (\ref{sec:Two-particle-Wess-Zumino-term}).
In the end, we provide the feasible physical realization schemes for third-order
topological insulators (TOTIs) and superconductors (TOTSCs), predicting the emergence
of zero corner modes in terms of surface Chern-Simons theory ($\tilde{\theta}$ characterization
and $\tilde{\mathrm{n}}_{\mathrm{WZ}}$ mechanism) (\ref{sec:Feasible physical realizations for TOTIs (TOTSCs)}).
The symmetry analysis for zero corner modes are also provided. 

\section{\label{sec:Chern-Simons-term-}Chern-Simons term $\tilde{\theta}$ characterization
for TOTIs (TOTSCs)}

\subsection{\label{subsec: Chern-Simons term} Chern-Simons term $\tilde{\theta}$ }

For 3D second-order topological insulators (SOTIs) or superconductors (SOTSCs), the
effective surface Hamiltonian is usually given by: $\mathcal{H}_{\mathrm{effective}}^{\mathrm{second}}=vk_{1}\Gamma_{1}+vk_{2}\Gamma_{2}+m_{1}\Gamma_{3}$
with $\boldsymbol{\Gamma}$ been gamma matrices, and the emergence of helical hinge
modes relies on sign inversion of effective mass $m_{1}$ of adjacent surfaces. In
order to induce the 3rd-order zero energy corner modes from 3D SOTIs or SOTSCs, we
also need to consider additional mass term $m_{2}\Gamma_{4}$ which can gap out this
helical hinge modes and change sign between adjacent hinges. Thus the effective surface
Hamiltonian for 3D TOTIs or TOTSCs can be written as : $\mathcal{H}_{\mathrm{effective}}^{\mathrm{3rd}}=vk_{1}\Gamma_{1}+vk_{2}\Gamma_{2}+m_{1}\Gamma_{3}+m_{2}\Gamma_{4}$,
where $\Gamma_{i}$ satisfies $\{\Gamma_{i},\Gamma_{j}\}=2\delta_{ij}$. Notice that
$\mathcal{H}_{\mathrm{effective}}=vk_{1}\Gamma_{1}+vk_{2}\Gamma_{2}$ represents
the Dirac (Majorana) cones of 3D TIs (TSCs) and ($m_{1},m_{2}$) can be viewed as
surface effective mass field. In the following, we will show that the Chern-Simons
term $\tilde{\theta}$ is related two winding numbers derived from surface mass fields
and Dirac (Majorana) cones of 3D TIs (TSCs). 

\begin{figure*}
\includegraphics{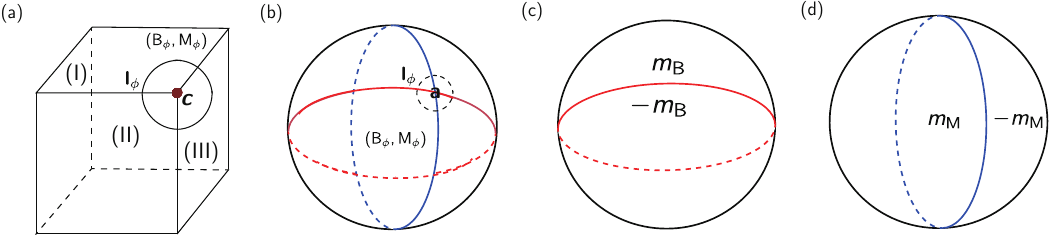}

\caption{\label{Fig1} Chern-Simons term $\tilde{\theta}$ characterization. (a) A schematic
for effective mass field $\mathrm{\boldsymbol{m}}=(\mathrm{B}_{\phi},\mathrm{M}_{\phi})$
in the surface of 3D TI (TSC). Here $(\mathrm{I})$, $(\mathrm{II})$, $(\mathrm{III})$
are three adjacent surfaces and corner $\boldsymbol{c}$ is a common point of these
surfaces. $\mathrm{\boldsymbol{l}}_{\phi}$ denotes a closed loop surrounding $\boldsymbol{c}$.
(b) We can deform such surface effective mass field to that in (b) without gap closing.
And the target corner $\boldsymbol{c}$ is mapped to a intersection $\mathrm{\boldsymbol{a}}$
of the blue and red lines which represent respectively the mass domain walls formed
by $m_{\mathrm{B}}$ in (c) and $m_{\mathrm{M}}$ in (d).}
\end{figure*}

As shown in Fig.\ref{Fig1}(a), we focus on a sample corner $\boldsymbol{c}$ that
is formed by three neighbouring surfaces $(\mathrm{I})$, $(\mathrm{II})$ and $(\mathrm{III})$
of 3D TIs (TSCs). For simplicity, we adopt the spherical coordinate frame. Hence,
the corresponding surface Dirac (Majorana) cone is captured by $\mathcal{H}_{\mathrm{effective}}=-\mathrm{k_{\mathrm{\varphi}}}\Gamma_{1}+\mathrm{k_{\theta}}\Gamma_{2}$,
where $\mathrm{k_{\mathrm{\varphi}}}$ and $\mathrm{k_{\theta}}$ are momentums.
Now we introduce the surface effective mass field ($m_{1},m_{2}$) to $\mathcal{H}_{\mathrm{effective}}$
and compute the Chern-Simons term $\tilde{\theta}$. It is well known that $\tilde{\theta}$
is defined on the manifold $k^{d}\times\mathcal{M}^{D}$ with $d+D=\mathrm{odd}$
\citep{SRyuRevModPhys2016-1}. To meet this need, in addition to the 2D momentum
space $(\mathrm{k}_{\varphi},\mathrm{k}_{\theta})$, we also focus on loop $\mathrm{\boldsymbol{l}}_{\phi}$
surrounding target corner $\boldsymbol{c}$ and take the real space parameter $\phi$
as a synthetic dimension of ring geometry $\mathcal{S}^{1}$. When mass field ($m_{1},m_{2}$)
is parameterized as ($\mathrm{B}_{\phi},\mathrm{M}_{\phi}$), the $\mathcal{H}_{\mathrm{effective}}^{\mathrm{3rd}}$
can be written down in a synthetic 3D space $k^{3}=k^{2}\times\mathcal{S}^{1}$ spanned
by $(\boldsymbol{k},\phi)$ with $\mathcal{H}_{\mathrm{effective}}^{\mathrm{3rd}}=-\mathrm{k_{\mathrm{\varphi}}}\Gamma_{1}+\mathrm{k_{\theta}}\Gamma_{2}+\mathrm{B}_{\phi}\Gamma_{3}+\mathrm{M}_{\phi}\Gamma_{4}$.
Thus the Chern-Simons term with respect to corner $\boldsymbol{c}$ is given by 
\begin{equation}
\tilde{\theta}=\frac{1}{4\pi}\int\mathrm{d}^{3}k\epsilon_{\mathrm{abc}}\mathrm{tr}[\mathcal{A}_{\mathrm{a}}\partial_{\mathrm{b}}\mathcal{A}_{\mathrm{c}}+\mathrm{i}\frac{2}{3}\mathcal{A}_{\mathrm{a}}\mathcal{A}_{\mathrm{b}}\mathcal{A}_{\mathrm{c}}],\label{eq:1-1}
\end{equation}
where $\mathcal{A}_{\mathrm{a}=(\boldsymbol{k},\phi),n,n^{\prime}}=-\mathrm{i}\tilde{\langle n|}\partial_{a}\tilde{|n^{\prime}\rangle}$
is non-Abelian Berry connection with $|\tilde{n}\rangle$ the eigenfunction for filled
degenerate bands $-E=-\sqrt{\mathrm{k_{\mathrm{\varphi}}^{2}}+\mathrm{k_{\theta}^{2}}+\mathrm{B}_{\phi}^{2}+\mathrm{M}_{\phi}^{2}}$.
In order to obtain the explicit form of $|\tilde{n}\rangle$, $\Gamma_{i}$ can be
taken as $\Gamma_{1}=\mu_{y}\gamma_{x}$, $\Gamma_{2}=-\mu_{y}\gamma_{z}$, $\Gamma_{3}=\mu_{y}\gamma_{y}$,
$\Gamma_{4}=\mu_{x}\gamma_{0}$. As a result, the degenerate eigenvectors are given
by $\tilde{|1\rangle}=\left(\begin{array}{cccc}
\frac{\mathrm{B}_{\phi}-\mathrm{i}\mathrm{k_{\mathrm{\varphi}}}}{\mathcal{N}} & \frac{-\mathrm{M}_{\phi}+\mathrm{i}\mathrm{\mathrm{k_{\theta}}}}{\mathcal{N}} & 0 & \frac{E}{\mathcal{N}}\end{array}\right)^{\mathrm{T}}$, $\tilde{|2\rangle}=\left(\begin{array}{cccc}
\frac{-\mathrm{M}_{\phi}-\mathrm{i}\mathrm{\mathrm{k_{\theta}}}}{\mathcal{N}} & \frac{-\mathrm{B}_{\phi}-\mathrm{i}\mathrm{k_{\mathrm{\varphi}}}}{\mathcal{N}} & \frac{E}{\mathcal{N}} & 0\end{array}\right)^{\mathrm{T}}$where $\mathcal{N}=\sqrt{2E^{2}}$. 
 Correspondingly, the non-Abelian Berry connections are easily obtained: 
\begin{align}
\mathcal{A}_{\phi}= & -\mathrm{i}\left(\begin{array}{cc}
\frac{(\mathrm{i}\mathrm{k_{\mathrm{\varphi}}})}{\mathcal{N}^{2}} & \frac{(\mathrm{M}_{\phi}+\mathrm{i}\mathrm{\mathrm{k_{\theta}}})}{\mathcal{N}^{2}}\\
\frac{(-\mathrm{M}_{\phi}+\mathrm{i}\mathrm{\mathrm{k_{\theta}}})}{\mathcal{N}^{2}} & \frac{(-\mathrm{i}\mathrm{k_{\mathrm{\varphi}})}}{\mathcal{N}^{2}}
\end{array}\right)\partial_{\phi}(\mathrm{B}_{\phi})-\mathrm{i}\left(\begin{array}{cc}
\frac{(\mathrm{i}\mathrm{\mathrm{k_{\theta}}})}{\mathcal{N}^{2}} & \frac{-(\mathrm{B}_{\phi}+\mathrm{i}\mathrm{k_{\mathrm{\varphi}})}}{\mathcal{N}^{2}}\\
\frac{(\mathrm{B}_{\phi}-\mathrm{i}\mathrm{k_{\mathrm{\varphi}})}}{\mathcal{N}^{2}} & \frac{(-\mathrm{i}\mathrm{\mathrm{k_{\theta})}}}{\mathcal{N}^{2}}
\end{array}\right)\partial_{\phi}(\mathrm{M}_{\phi})\nonumber \\
\mathcal{A}_{\mathrm{\boldsymbol{k}}}= & -\mathrm{i}\left(\begin{array}{cc}
\frac{-\mathrm{i}\mathrm{B}_{\phi}}{\mathcal{N}^{2}} & \frac{\mathrm{i}\mathrm{M}_{\phi}-\mathrm{\mathrm{k_{\theta}}}}{\mathcal{N}^{2}}\\
\frac{\mathrm{i}\mathrm{M}_{\phi}+\mathrm{\mathrm{k_{\theta}}}}{\mathcal{N}^{2}} & \frac{\mathrm{i}\mathrm{B}_{\phi}}{\mathcal{N}^{2}}
\end{array}\right)\nabla_{\mathrm{\boldsymbol{k}}}(\mathrm{k_{\mathrm{\varphi}})}-\mathrm{i}\left(\begin{array}{cc}
\frac{-\mathrm{i}\mathrm{M}_{\phi}}{\mathcal{N}^{2}} & \frac{-\mathrm{i}\mathrm{B}_{\phi}+\mathrm{k_{\mathrm{\varphi}}}}{\mathcal{N}^{2}}\\
\frac{-\mathrm{i}\mathrm{B}_{\phi}-\mathrm{k_{\mathrm{\varphi}}}}{\mathcal{N}^{2}} & \frac{\mathrm{\mathrm{i}\mathrm{M}_{\phi}}}{\mathcal{N}^{2}}
\end{array}\right)\nabla_{\mathrm{\boldsymbol{k}}}(\mathrm{\mathrm{k_{\theta}})}.
\end{align}
And the explicit form of $\tilde{\theta}$ reads $\frac{1}{4\pi}\int\mathrm{d}^{3}k\mathrm{tr}\{[\mathcal{A}_{\mathrm{k_{\mathrm{\varphi}}}}(\partial_{\mathrm{k_{\theta}}}\mathcal{A}_{\phi}-\partial_{\phi}\mathcal{A}_{\mathrm{k_{\theta}}})+\mathcal{A}_{\mathrm{k_{\theta}}}(\partial_{\phi}\mathcal{A}_{\mathrm{k_{\mathrm{\varphi}}}}-\partial_{\mathrm{k_{\mathrm{\varphi}}}}\mathcal{A}_{\phi})+\mathcal{A}_{\phi}(\partial_{\mathrm{k_{\mathrm{\varphi}}}}\mathcal{A}_{\mathrm{k_{\theta}}}-\partial_{\mathrm{k_{\theta}}}\mathcal{A}_{\mathrm{k_{\mathrm{\varphi}}}})]+\mathrm{i}\frac{2}{3}[\mathcal{A}_{\mathrm{k_{\mathrm{\varphi}}}}(\mathcal{A}_{\mathrm{k_{\theta}}}\mathcal{A}_{\phi}-\mathcal{A}_{\phi}\mathcal{A}_{\mathrm{k_{\theta}}})+\mathcal{A}_{\mathrm{\mathrm{k_{\theta}}}}(\mathcal{A}_{\phi}\mathcal{A}_{\mathrm{k_{\mathrm{\varphi}}}}-\mathcal{A}_{\mathrm{k_{\mathrm{\varphi}}}}\mathcal{A}_{\phi})+\mathcal{A}_{\phi}(\mathcal{A}_{\mathrm{k_{\mathrm{\varphi}}}}\mathcal{A}_{\mathrm{k_{\theta}}}-\mathcal{A}_{\mathrm{k_{\theta}}}\mathcal{A}_{k_{\mathrm{\varphi}}})]\}$.
To simplify the computation, we define four matrices $\mathcal{A}_{\mathrm{B}_{\phi}}$,
$\mathcal{A}_{\mathrm{M}_{\phi}}$, $\mathcal{A}_{\varphi}$ and $\mathcal{A}_{\theta}$
which take the form of : 
\begin{align}
\mathcal{A}_{\mathrm{B}_{\phi}}=\left(\begin{array}{cc}
\frac{(\mathrm{i}\mathrm{k_{\mathrm{\varphi}}})}{\mathcal{N}^{2}} & \frac{(\mathrm{M}_{\phi}+\mathrm{i}\mathrm{\mathrm{k_{\theta}}})}{\mathcal{N}^{2}}\\
\frac{(-\mathrm{M}_{\phi}+\mathrm{i}\mathrm{\mathrm{k_{\theta}}})}{\mathcal{N}^{2}} & \frac{(-\mathrm{i}\mathrm{k_{\mathrm{\varphi}})}}{\mathcal{N}^{2}}
\end{array}\right) & ;\quad\mathcal{A}_{\mathrm{M}_{\phi}}=\left(\begin{array}{cc}
\frac{(\mathrm{i}\mathrm{\mathrm{k_{\theta}}})}{\mathcal{N}^{2}} & \frac{-(\mathrm{B}_{\phi}+\mathrm{i}\mathrm{k_{\mathrm{\varphi}})}}{\mathcal{N}^{2}}\\
\frac{(\mathrm{B}_{\phi}-\mathrm{i}\mathrm{k_{\mathrm{\varphi}})}}{\mathcal{N}^{2}} & \frac{(-\mathrm{i}\mathrm{\mathrm{k_{\theta})}}}{\mathcal{N}^{2}}
\end{array}\right)\nonumber \\
\mathcal{A}_{\varphi}=\left(\begin{array}{cc}
\frac{-\mathrm{i}\mathrm{B}_{\phi}}{\mathcal{N}^{2}} & \frac{\mathrm{i}\mathrm{M}_{\phi}-\mathrm{\mathrm{k_{\theta}}}}{\mathcal{N}^{2}}\\
\frac{\mathrm{i}\mathrm{M}_{\phi}+\mathrm{\mathrm{k_{\theta}}}}{\mathcal{N}^{2}} & \frac{\mathrm{i}\mathrm{B}_{\phi}}{\mathcal{N}^{2}}
\end{array}\right) & ;\quad\mathcal{A}_{\theta}=\left(\begin{array}{cc}
\frac{-\mathrm{i}\mathrm{M}_{\phi}}{\mathcal{N}^{2}} & \frac{-\mathrm{i}\mathrm{B}_{\phi}+\mathrm{k_{\mathrm{\varphi}}}}{\mathcal{N}^{2}}\\
\frac{-\mathrm{i}\mathrm{B}_{\phi}-\mathrm{k_{\mathrm{\varphi}}}}{\mathcal{N}^{2}} & \frac{\mathrm{\mathrm{i}\mathrm{M}_{\phi}}}{\mathcal{N}^{2}}
\end{array}\right).\label{eq:3-1}
\end{align}
So the non-Abelian Berry connections are formulated as $\mathcal{A}_{k_{\mathrm{\varphi}}}=-\mathrm{i}\mathcal{A}_{\varphi}$,
$\mathcal{A}_{k_{\theta}}=-\mathrm{i}\mathcal{A}_{\theta}$ and $\mathcal{A}_{\phi}=-\mathrm{i}\mathcal{A}_{\mathrm{B}_{\phi}}\partial_{\phi}(\mathrm{B}_{\phi})-\mathrm{i}\mathcal{A}_{\mathrm{M}_{\phi}}\partial_{\phi}(\mathrm{M}_{\phi})$.
Substituting them into the explicit form of $\tilde{\theta}$, we reach that $\mathrm{\epsilon_{\mathrm{abc}}tr}[\mathcal{A}_{\mathrm{a}}\partial_{\mathrm{b}}\mathcal{A}_{\mathrm{c}}+\mathrm{i}\frac{2}{3}\mathcal{A}_{\mathrm{a}}\mathcal{A}_{\mathrm{b}}\mathcal{A}_{\mathrm{c}}]=2\frac{\mathrm{M}_{\phi}\partial_{\phi}(\mathrm{B}_{\phi})-\mathrm{B}_{\phi}\partial_{\phi}(\mathrm{M}_{\phi})}{E^{4}}$
. Finally, the Chern-Simons term $\tilde{\theta}$ can be expressed in a concise
form: 
\begin{align}
\tilde{\theta}= & \frac{1}{2\pi}\int_{-\infty}^{+\infty}\int_{-\infty}^{+\infty}\oint_{\mathrm{l}_{\boldsymbol{\phi}}}\mathrm{d}\mathrm{k_{\mathrm{\varphi}}\mathrm{d}\mathrm{k_{\mathrm{\theta}}}\mathrm{d}\phi}\frac{\mathrm{M}_{\phi}\partial_{\phi}(\mathrm{B}_{\phi})-\mathrm{B}_{\phi}\partial_{\phi}(\mathrm{M}_{\phi})}{(\mathrm{k_{\mathrm{\varphi}}^{2}}+\mathrm{k_{\mathrm{\theta}}^{2}+\mathrm{B}_{\phi}^{2}+\mathrm{M}_{\phi}^{2}})^{2}}=\pi\frac{1}{2\pi}\oint_{\mathrm{l}_{\boldsymbol{\phi}}}\mathrm{d}\phi\frac{\mathrm{M}_{\phi}\partial_{\phi}(\mathrm{B}_{\phi})-\mathrm{B}_{\phi}\partial_{\phi}(\mathrm{M}_{\phi})}{\mathrm{B}_{\phi}^{2}+\mathrm{M}_{\phi}^{2}}=\pi\mathcal{W}_{\boldsymbol{\mathrm{m}}},\label{eq:4-1}
\end{align}
where $\boldsymbol{\mathrm{m}}=(\mathrm{B}_{\phi},\mathrm{M}_{\phi})$ is the effective
mass field defined on the surface of TOTIs (TOTSCs). Eq.(\ref{eq:4-1}) shows that
Chern-Simons term $\tilde{\theta}$ is associated with the winding number of surface
effective mass field $\boldsymbol{\mathrm{m}}$. In fact, $\tilde{\theta}$ is also
dependent on the surface Dirac (Majorana) cone winding number in $\boldsymbol{k}$
space besides the $\mathcal{W}_{\boldsymbol{\mathrm{m}}}$. To demonstrate this we
consider a more general form: $\mathcal{H}_{\mathrm{eff}}^{\prime}=\mathrm{h}_{1}(\mathrm{k}_{\varphi},\mathrm{k}_{\theta})\mu_{y}\gamma_{x}+\mathrm{h}_{2}(\mathrm{k}_{\varphi},\mathrm{k}_{\theta})\mu_{y}\gamma_{z}+\mathrm{B}_{\phi}\mu_{y}\gamma_{y}+\mathrm{M}_{\phi}\mu_{x}\gamma_{0}$
where $\mathrm{\boldsymbol{h}}=(\mathrm{h}_{1},\mathrm{h}_{2})$ is a linear function
of $\boldsymbol{k}$ (e.g. $\mathrm{h}_{1}=-\mathrm{k_{\mathrm{\varphi}}}$, $\mathrm{h}_{2}=-\mathrm{k_{\theta}}$).
Thus the corresponding degenerate ground states for $\mathcal{H}_{\mathrm{eff}}^{\prime}$
become $|\tilde{\mathrm{I}}\rangle=\left(\begin{array}{cccc}
\frac{\mathrm{B}_{\phi}+\mathrm{i}\mathrm{h}_{1}}{\mathcal{N}} & \frac{-\mathrm{M}_{\phi}+\mathrm{i}\mathrm{h}_{2}}{\mathcal{N}} & 0 & \frac{E}{\mathcal{N}}\end{array}\right)^{\mathrm{T}}$,$|\tilde{\mathrm{II}}\rangle=\left(\begin{array}{cccc}
\frac{-\mathrm{M}_{\phi}-\mathrm{i}\mathrm{h}_{2}}{\mathcal{N}} & \frac{-\mathrm{B}_{\phi}+\mathrm{i}\mathrm{h}_{1}}{\mathcal{N}} & \frac{E}{\mathcal{N}} & 0\end{array}\right)^{\mathrm{T}}$ with $E=\sqrt{\mathrm{h}_{1}^{2}+\mathrm{h}_{2}^{2}+\mathrm{B}_{\phi}^{2}+\mathrm{M}_{\phi}^{2}}$
and $\mathcal{N}=\sqrt{2E^{2}}$. Similarly, the non-Abelian Berry connections are
given by $\mathcal{A}_{\phi}=-\mathrm{i}\mathcal{A}_{\mathrm{B}_{\phi}}\partial_{\phi}(\mathrm{B}_{\phi})-\mathrm{i}\mathcal{A}_{\mathrm{M}_{\phi}}\partial_{\phi}(\mathrm{M}_{\phi})$,
$\mathcal{A}_{k_{\mathrm{\varphi}}}=-\mathrm{i}\mathcal{A}_{1}\partial_{\mathrm{k}_{\varphi}}(\mathrm{h}_{1}\mathrm{)}-\mathrm{i}\mathcal{A}_{2}\partial_{\mathrm{k}_{\varphi}}(\mathrm{\mathrm{h}_{2})}$
, $\mathcal{A}_{k_{\theta}}=-\mathrm{i}\mathcal{A}_{1}\partial_{\mathrm{k}_{\theta}}(\mathrm{h}_{1}\mathrm{)}-\mathrm{i}\mathcal{A}_{2}\partial_{\mathrm{k}_{\theta}}(\mathrm{\mathrm{h}_{2})}$.
And $\mathcal{A}_{\mathrm{B}_{\phi}}$, $\mathcal{A}_{\mathrm{M}_{\phi}}$, $\mathcal{A}_{1}$
, $\mathcal{A}_{2}$ read 
\begin{align}
\mathcal{A}_{\mathrm{B}_{\phi}}=\left(\begin{array}{cc}
\frac{(-\mathrm{i}\mathrm{h}_{1})}{\mathcal{N}^{2}} & \frac{(\mathrm{M}_{\phi}+\mathrm{i}\mathrm{h}_{2})}{\mathcal{N}^{2}}\\
\frac{(-\mathrm{M}_{\phi}+\mathrm{i}\mathrm{h}_{2})}{\mathcal{N}^{2}} & \frac{(\mathrm{i}\mathrm{h}_{1})}{\mathcal{N}^{2}}
\end{array}\right); & \quad\mathcal{A}_{\mathrm{M}_{\phi}}=\left(\begin{array}{cc}
\frac{(\mathrm{i}\mathrm{h}_{2})}{\mathcal{N}^{2}} & \frac{(-\mathrm{B}_{\phi}+\mathrm{i}\mathrm{h}_{1})}{\mathcal{N}^{2}}\\
\frac{(\mathrm{B}_{\phi}+\mathrm{i}\mathrm{h}_{1})}{\mathcal{N}^{2}} & \frac{(-\mathrm{\mathrm{i\mathrm{h}_{2})}}}{\mathcal{N}^{2}}
\end{array}\right)\nonumber \\
\mathcal{A}_{1}=\left(\begin{array}{cc}
\frac{(\mathrm{i}\mathrm{B}_{\phi})}{\mathcal{N}^{2}} & \frac{(-\mathrm{i}\mathrm{M}_{\phi}+\mathrm{h}_{2})}{\mathcal{N}^{2}}\\
\frac{(-\mathrm{i}\mathrm{M}_{\phi}-\mathrm{\mathrm{h}_{2})}}{\mathcal{N}^{2}} & \frac{-\mathrm{i}\mathrm{B}_{\phi}}{\mathcal{N}^{2}}
\end{array}\right); & \quad\mathcal{A}_{2}=\left(\begin{array}{cc}
\frac{(-\mathrm{i}\mathrm{M}_{\phi})}{\mathcal{N}^{2}} & \frac{(-\mathrm{i}\mathrm{B}_{\phi}-\mathrm{\mathrm{h}_{1})}}{\mathcal{N}^{2}}\\
\frac{(-\mathrm{i}\mathrm{B}_{\phi}+\mathrm{\mathrm{h}_{1})}}{\mathcal{N}^{2}} & \frac{\mathrm{i}\mathrm{\mathrm{M}_{\phi}}}{\mathcal{N}^{2}}
\end{array}\right).
\end{align}
After some algebraic operation, we obtain $\mathrm{\epsilon_{\mathrm{abc}}tr}[\mathcal{A}_{\mathrm{a}}\partial_{\mathrm{b}}\mathcal{A}_{\mathrm{c}}+\mathrm{i}\frac{2}{3}\mathcal{A}_{\mathrm{a}}\mathcal{A}_{\mathrm{b}}\mathcal{A}_{\mathrm{c}}]=\frac{2[\partial_{\mathrm{k}_{\varphi}}(\mathrm{h}_{1}\mathrm{)}\partial_{\mathrm{k}_{\theta}}(\mathrm{\mathrm{h}_{2})}-\partial_{\mathrm{k}_{\theta}}(\mathrm{h}_{1}\mathrm{)}\partial_{\mathrm{k}_{\varphi}}(\mathrm{\mathrm{h}_{2})}][\mathrm{M}_{\phi}\partial_{\phi}(\mathrm{B}_{\phi})-\mathrm{B}_{\phi}\partial_{\phi}(\mathrm{M}_{\phi})]}{E^{4}}$.
Then the Chern-Simons term $\tilde{\theta}$ takes the form of 
\begin{align}
\tilde{\theta}= & \frac{1}{2\pi}\oint_{\mathrm{l}_{\boldsymbol{\phi}}}\mathrm{d}\phi[\frac{\mathrm{M}_{\phi}\partial_{\phi}(\mathrm{B}_{\phi})-\mathrm{B}_{\phi}\partial_{\phi}(\mathrm{M}_{\phi})}{\mathrm{B}_{\phi}^{2}+\mathrm{M}_{\phi}^{2}}]\int_{-\infty}^{+\infty}\int_{-\infty}^{+\infty}\mathrm{d}\mathrm{k}_{\varphi}\mathrm{d}\mathrm{k}_{\theta}\frac{[\partial_{\mathrm{k}_{\varphi}}(\mathrm{h}_{1}\mathrm{)}\partial_{\mathrm{k}_{\theta}}(\mathrm{\mathrm{h}_{2})}-\partial_{\mathrm{k}_{\theta}}(\mathrm{h}_{1}\mathrm{)}\partial_{\mathrm{k}_{\varphi}}(\mathrm{\mathrm{h}_{2})}](\mathrm{B}_{\phi}^{2}+\mathrm{M}_{\phi}^{2})}{E^{4}}\nonumber \\
= & \mathcal{W}_{\mathrm{\boldsymbol{m}}}\int_{\mathrm{\boldsymbol{k}\hspace{0.3em}space}}\frac{\mathrm{d}^{2}\boldsymbol{\mathrm{k}}\cdot(\nabla_{\mathrm{\boldsymbol{k}}}\mathrm{h}_{1}\times\nabla_{\mathrm{\boldsymbol{k}}}\mathrm{h}_{2})(\mathrm{B}_{\phi}^{2}+\mathrm{M}_{\phi}^{2})}{(\mathrm{h}_{1}^{2}+\mathrm{h}_{2}^{2}+\mathrm{B}_{\phi}^{2}+\mathrm{M}_{\phi}^{2})^{2}}.\label{eq:6}
\end{align}
Both $\mathrm{h}_{1}$ and $\mathrm{h}_{2}$ depend linearly on $\mathrm{\boldsymbol{k}}$,
so $\nabla_{\mathrm{\boldsymbol{k}}}\mathrm{h}_{1}(\mathrm{\boldsymbol{k}})\times\nabla_{\mathrm{\boldsymbol{k}}}\mathrm{h}_{2}(\mathrm{\boldsymbol{k}})$
is a constant. And we have $\int_{\mathrm{\boldsymbol{k}\hspace{0.3em}space}}\frac{\mathrm{d}^{2}\boldsymbol{\mathrm{k}}\cdot(\nabla_{\mathrm{\boldsymbol{k}}}\mathrm{h}_{1}\times\nabla_{\mathrm{\boldsymbol{k}}}\mathrm{h}_{2})(\mathrm{B}_{\phi}^{2}+\mathrm{M}_{\phi}^{2})}{(\mathrm{h}_{1}^{2}+\mathrm{h}_{2}^{2}+\mathrm{B}_{\phi}^{2}+\mathrm{M}_{\phi}^{2})^{2}}=\int_{\mathrm{\boldsymbol{k}\hspace{0.3em}space}-\mathcal{\boldsymbol{S}}}\frac{\mathrm{d}^{2}\boldsymbol{\mathrm{k}}\cdot(\nabla_{\mathrm{\boldsymbol{k}}}\mathrm{h}_{1}\times\nabla_{\mathrm{\boldsymbol{k}}}\mathrm{h}_{2})(\mathrm{B}_{\phi}^{2}+\mathrm{M}_{\phi}^{2})}{(\mathrm{h}_{1}^{2}+\mathrm{h}_{2}^{2})^{2}}$
where \textbf{$\mathcal{\boldsymbol{S}}$} denotes an area $\{\mathrm{\boldsymbol{k}}|\mathrm{h}_{1}^{2}(\mathrm{\boldsymbol{k}})+\mathrm{h}_{2}^{2}(\mathrm{\boldsymbol{k}})\leq\mathrm{B}_{\phi}^{2}+\mathrm{M}_{\phi}^{2}\}$
in $\mathrm{\boldsymbol{k}}$ space and its boundary $\partial\boldsymbol{S}$ is
a closed loop along which $\mathrm{h}_{1}^{2}(\mathrm{\boldsymbol{k}})+\mathrm{h}_{2}^{2}(\mathrm{\boldsymbol{k}})=\mathrm{B}_{\phi}^{2}+\mathrm{M}_{\phi}^{2}$.
In other word, $\partial\boldsymbol{S}$ resembles a ``Fermi surface'' which is
obtained by crossing the Dirac (Majorana) cone with ``chemical potential'' $\mu=\sqrt{\mathrm{B}_{\phi}^{2}+\mathrm{M}_{\phi}^{2}}$
(Fig.1(b) in main text). It is well known that $\nabla_{\mathrm{\boldsymbol{k}}}\mathrm{h}_{1}\times\nabla_{\mathrm{\boldsymbol{k}}}\mathrm{h}_{2}=\nabla_{\mathrm{\boldsymbol{k}}}\times(\mathrm{h}_{1}\nabla_{\mathrm{\boldsymbol{k}}}\mathrm{h}_{2}-\mathrm{h}_{2}\nabla_{\mathrm{\boldsymbol{k}}}\mathrm{h}_{1})$,
then Eq.(\ref{eq:6}) becomes $\tilde{\theta}=\mathcal{W}_{\mathrm{\boldsymbol{m}}}\int_{\mathrm{\boldsymbol{k}\hspace{0.3em}space}-\mathcal{\boldsymbol{S}}}\frac{\mathrm{d}^{2}\boldsymbol{\mathrm{k}}\cdot[\nabla_{\mathrm{\boldsymbol{k}}}\times(\mathrm{h}_{1}\nabla_{\mathrm{\boldsymbol{k}}}\mathrm{h}_{2}-\mathrm{h}_{2}\nabla_{\mathrm{\boldsymbol{k}}}\mathrm{h}_{1})](\mathrm{B}_{\phi}^{2}+\mathrm{M}_{\phi}^{2})}{(\mathrm{h}_{1}^{2}+\mathrm{h}_{2}^{2})^{2}}$.
One can readily check that $\text{\ensuremath{\nabla_{\mathrm{\boldsymbol{k}}}\times}[\ensuremath{\frac{\mathrm{h}_{2}\nabla_{\mathrm{\boldsymbol{k}}}\mathrm{h}_{1}-\mathrm{h}_{1}\nabla_{\mathrm{\boldsymbol{k}}}\mathrm{h}_{2}}{2(\mathrm{h}_{1}^{2}+\mathrm{h}_{2}^{2})^{2}}}]=\ensuremath{\frac{[\nabla_{\mathrm{\boldsymbol{k}}}\times(\mathrm{h}_{1}\nabla_{\mathrm{\boldsymbol{k}}}\mathrm{h}_{2}-\mathrm{h}_{2}\nabla_{\mathrm{\boldsymbol{k}}}\mathrm{h}_{1})]}{(\mathrm{h}_{1}^{2}+\mathrm{h}_{2}^{2})^{2}}}.}$
Hence 
\begin{equation}
\tilde{\theta}=\mathcal{W}_{\mathrm{\boldsymbol{m}}}\int_{\mathrm{\boldsymbol{k}\hspace{0.3em}space}-\mathcal{\boldsymbol{S}}}\mathrm{d}^{2}\boldsymbol{\mathrm{k}}\cdot\{\nabla_{\mathrm{\boldsymbol{k}}}\times[\frac{\mathrm{h}_{2}\nabla_{\mathrm{\boldsymbol{k}}}\mathrm{h}_{1}-\mathrm{h}_{1}\nabla_{\mathrm{\boldsymbol{k}}}\mathrm{h}_{2}}{2(\mathrm{h}_{1}^{2}+\mathrm{h}_{2}^{2})^{2}}]\}(\mathrm{B}_{\phi}^{2}+\mathrm{M}_{\phi}^{2}).
\end{equation}
Using the Stokes\textquoteright{} theorem, we obtain 
\begin{align}
\tilde{\theta} & =\pi\mathcal{W}_{\mathrm{\boldsymbol{m}}}\frac{1}{2\pi}\oint_{\partial\boldsymbol{S}}\frac{\mathrm{d}\mathrm{\boldsymbol{l}}\cdot(\mathrm{h}_{2}\nabla_{\mathrm{\boldsymbol{k}}}\mathrm{h}_{1}-\mathrm{h}_{1}\nabla_{\mathrm{\boldsymbol{k}}}\mathrm{h}_{2})}{(\mathrm{h}_{2}^{2}+\mathrm{h}_{1}^{2})}=\pi\mathcal{W}_{\mathrm{\boldsymbol{h}}}\mathcal{W}_{\mathrm{\boldsymbol{m}}},\label{eq:8}
\end{align}
where $\mathcal{W}_{\mathrm{\boldsymbol{h}}}$ denotes the Dirac (Majorana) cone
winding number in $\boldsymbol{\mathrm{k}}$ space. So far, we have shown that the
CS term $\tilde{\theta}$, which is defined within a synthetic 3D space $(\boldsymbol{k},\phi)$,
relies on two winding numbers derived from surface mass fields and Dirac (Majorana)
cones. Furthermore, Eq.(\ref{eq:8}) also shows that Chern-Simons term can reduce
to the Hopf invariant which captures the linking number of the inverse images of
two points in the target space $\boldsymbol{S}^{2}$ of effective surface Hamiltonian
$\mathcal{H}_{\mathrm{effective}}(\mathrm{\boldsymbol{k}},\phi)$ \citep{CChanPhysRevLett2017-1}.
In general, for Dirac (Majorana) cone dictated by the Hamiltonian $\mathrm{\boldsymbol{h}}\cdot\boldsymbol{\Gamma}$
(e.g. $\mathrm{h}_{1}=-\mathrm{k_{\mathrm{\varphi}}}$, $\mathrm{h}=\mathrm{k_{\theta}}$),
$\mathcal{W}_{\mathrm{\boldsymbol{h}}}=1$, so $\tilde{\theta}$ returns to Eq.(\ref{eq:4-1}). 

\subsection{\textcolor{red}{\label{par:The relation between non-zero =00005Ctilde=00007B=00005Ctheta=00007D and zero corner mode }}The
relation between nonzero $\tilde{\theta}$ and zero corner mode }

In the previous discussion, we have got the Chern-Simons term which reads $\tilde{\theta}=\pi\mathcal{W}_{\mathrm{\boldsymbol{h}}}\mathcal{W}_{\mathrm{\boldsymbol{m}}}$.
Now we will demonstrate that a nonzero $\tilde{\theta}$ can correspond to a zero
corner mode. For convenience, we deform the cubic sample in Fig.\ref{Fig1}(a) to
a sphere shown in Fig.\ref{Fig1}(b) without gap closing, thus target corner $\mathrm{\boldsymbol{c}}$
is corrrespondingly mapped to the intersection $\mathrm{a}$. Now we first focus
on sphere sample and consider the following effective Hamiltonian: 
\begin{equation}
\mathcal{H}_{\mathrm{eff}}=-\mathrm{k_{\mathrm{\varphi}}}\mu_{y}\gamma_{x}-\mathrm{k_{\theta}}\mu_{y}\gamma_{z}+m_{\mathrm{B}}(\theta)\mu_{x}\gamma_{0}-m_{\mathrm{M}}(\varphi)\mu_{y}\gamma_{y}.\label{eq:9}
\end{equation}
According to the discussion in \citep{BenalcazarScience2017-1,BenalcazarPRB2017-1},
we can obtain the zero-energy localized solution. As plotted in Fig.\ref{Fig1}(c),
we will treat the $\theta$-edge as a domain wall where $m_{\mathrm{B}}$ steps from
positive (inside the topological phase) to negative (outside the topological phase):
$m_{\mathrm{B}}(\theta)>0\hspace{0.5em}(\mathrm{topological},\theta<\theta_{0})$,
$m_{\mathrm{B}}(\theta)<0\hspace{0.5em}(\mathrm{trivial},\theta_{0}>\theta)$ and
find the equation $(\mathrm{i}\partial_{\theta}\mu_{y}\gamma_{z}+m_{\mathrm{B}}(\theta)\mu_{x}\gamma_{0})\Psi(\theta)=0$.
With the ansatz $\Psi(\theta)=\mathrm{exp}(\int_{\theta_{0}}^{\theta}m_{\mathrm{B}}(\theta^{\prime})d\theta^{\prime})\Phi_{\theta}$
for some constant spinor $\Phi_{\theta}$, the matrix equation can be simplified
to $(\mathrm{I}-\mu_{z}\otimes\gamma_{z})\Phi_{x}=0$ and two solutions $\Phi_{\theta1}=(1,0,0,0)^{\mathrm{T}}$,
$\Phi_{\theta2}=(0,0,0,1)^{\mathrm{T}}$ which correspond to the positive eigenstate
of $\mu_{z}\otimes\gamma_{z}$ are obtained. Then we project the rest of the Hamiltonian
($\mathcal{H}_{\mathrm{rest}}=-\mathrm{k_{\mathrm{\varphi}}}\mu_{y}\gamma_{x}-m_{\mathrm{M}}(\varphi)\mu_{y}\gamma_{y}$)
in Eq.(\ref{eq:9}) onto the subspace spanned by $|\Phi_{\theta1}\rangle$ and $|\Phi_{\theta2}\rangle$
to find the edge Hamiltonian $\mathcal{H}_{\mathrm{edge},\theta}$ 
\begin{align}
\mathcal{H}_{\mathrm{edge},\theta} & =\left(\begin{array}{cc}
\langle\Phi_{\theta1}|\mathcal{H}_{\mathrm{rest}}|\Phi_{\theta1}\rangle & \langle\Phi_{\theta1}|\mathcal{H}_{\mathrm{rest}}|\Phi_{\theta2}\rangle\\
\langle\Phi_{\theta2}|\mathcal{H}_{\mathrm{rest}}|\Phi_{\theta1}\rangle & \langle\Phi_{\theta2}|\mathcal{H}_{\mathrm{rest}}|\Phi_{\theta2}\rangle
\end{array}\right)=-k_{\varphi}\tau_{y}+m_{\mathrm{M}}(\varphi)\tau_{x}.
\end{align}
Similarly, we also reach $\mathcal{H}_{\mathrm{edge},\varphi}=-k_{\theta}\nu_{y}+m_{\mathrm{B}}(\theta)\nu_{x}$
{[}Fig.\ref{Fig1}(d){]}. Now, we can get the simultaneous zero mode by considering
a intersection, i.e., either an $\theta$-edge with a $\varphi$ domain wall or a
$\varphi$-edge with an $\theta$ domain wall. Further, we find the following two
matrix equations for such two configurations respectively: $(\mathrm{I}-\tilde{\tau_{z}})\phi_{\theta,\varphi}=0$
and $(\mathrm{I}-\tilde{\nu_{z}})\phi_{\varphi,\theta}=0$. As a result, the zero
mode $\Phi=\left(\begin{array}{cc}
1 & 0\end{array}\right)$ for an $\theta$-edge with a $\varphi$ domain wall is the positive eigenstate of
$\tilde{\tau_{z}}$ while that for the $\varphi$-edge with a $\theta$ domain wall
is the positive eigenstate of $\tilde{\nu_{z}}$. That is to say, the zero mode is
a simultaneous zero mode of $\mathcal{H}_{\mathrm{edge},\theta}$ and $\mathcal{H}_{\mathrm{edge},\varphi}$,
which indicates a single mode appears at the intersection. 

We have illustrated that a zero mode emerges at the intersection of two domain walls
under the mass configurations considered above: $m_{\mathrm{B}}(\theta)>0\hspace{0.5em}(\mathrm{topological},\theta<\theta_{0})$
; $m_{\mathrm{B}}(\theta)<0\hspace{0.5em}(\mathrm{trivial},\theta_{0}>\theta)$ and
$m_{\mathrm{M}}(\varphi)>0\hspace{0.5em}(\mathrm{topological},\varphi<\varphi_{0})$
; $m_{\mathrm{M}}(\varphi)<0\hspace{0.5em}(\mathrm{trivial},\varphi_{0}>\varphi)$.
Next, we compute the Chern-Simons term $\tilde{\theta}$. Since $\mathcal{H}_{\mathrm{eff}}^{\mathrm{h}}=-\mathrm{k_{\mathrm{\varphi}}}\mu_{y}\gamma_{x}-\mathrm{k_{\theta}}\mu_{y}\gamma_{z}$
is a Dirac-type Hamiltonian, we can readily show that $\mathcal{W}_{\mathrm{\boldsymbol{h}}}=1$.
Additionally, to obtain the winding number $\mathcal{W}_{\mathrm{\boldsymbol{m}}}$
of such effective mass field $\boldsymbol{\mathrm{m}}$, $\phi\in[0,2\pi)$ {[}Fig.\ref{Fig1}(b){]}
is adopted to parameterize the $\boldsymbol{\mathrm{m}}$. Specifically, $m_{\mathrm{B}}(\theta)\Rightarrow m_{\mathrm{B}}\cos\phi$
and $m_{\mathrm{M}}(\varphi)\Rightarrow m_{\mathrm{M}}\sin\phi$, where $m_{\mathrm{B}}=m_{\mathrm{M}}=m$.
According to Eq.(\ref{eq:4-1}), the Chern-Simons term is computed: 
\begin{align}
\tilde{\theta} & =\pi\mathcal{W}_{\mathrm{\boldsymbol{h}}}\mathcal{W}_{\mathrm{\boldsymbol{m}}}=\pi[\frac{1}{2\pi}\int_{0}^{2\pi}\mathrm{d}\phi\frac{-m^{2}\sin\phi\partial_{\phi}(\cos\phi)+m^{2}\cos\phi\partial_{\phi}(\sin\phi)}{m^{2}}]=\frac{1}{2}\int_{0}^{2\pi}\mathrm{d}\phi=\pi
\end{align}
In contrast, we also calculate the $\tilde{\theta}$ for a different effective mass
field $\mathrm{\boldsymbol{m}^{\prime}}$ which can't enable the emergence of a zero
mode: $m_{\mathrm{B}}(\theta)\Rightarrow-m_{\mathrm{B}}\cos\phi$ and $m_{\mathrm{M}}(\varphi)\Rightarrow-m_{\mathrm{M}}\cos\phi$.
Hence we obtain 
\begin{align}
\tilde{\theta}^{\prime} & =\pi\mathcal{W}_{\mathrm{\boldsymbol{h}}}^{\prime}\mathcal{W}_{\mathrm{\boldsymbol{m}}}^{\prime}=\pi[\frac{1}{2\pi}\int_{0}^{2\pi}\mathrm{d}\phi\frac{m^{2}\cos\phi\partial_{\phi}(\cos\phi)-m^{2}\cos\phi\partial_{\phi}(\cos\phi)}{m^{2}}]=0
\end{align}
Till now, we have shown that the existence of zero mode at intersection of domain
walls is related to the Chern-Simons term $\tilde{\theta}$. Owing to the fact that
the cube and sphere are topologically equivalent and the corner $\mathrm{\boldsymbol{c}}$
is mapped to the intersection $\mathrm{a}$ after such deformation, a zero mode at
$\mathrm{a}$ must correspond to a zero mode at corner $\mathrm{\boldsymbol{c}}$,
leading to the TOTIs (TOTSCs). In other word, a topologically protected zero corner
mode is characterized by a nonzero Chern-Simons term ($\tilde{\theta}=\pi$). And
since the effective surface Hamiltonian of TOTIs (TOTSCs) resembles that of the 2D
$p\pm ip$ superconductors with Majorana zero mode at the vortex core, it belongs
to the class D of tenfold classification. Finally, it is noteworthy that the characterization
theoy based on $\tilde{\theta}$ is built on the general surface effective Hamiltonian,
rather than the specific detailed bulk properties. Consequently, this approach is
applicable to all TOTIs and TOTSCs, irrespective of the methods used to achieve these
topological phases. In conclusion, it stands as a robust characterization theory
that unifies the understanding of zero corner modes in both TOTIs and TOTSCs. 

\section{\label{sec:Two-particle-Wess-Zumino-term}Two-particle Wess-Zumino term $\tilde{\mathrm{n}}_{\mathrm{WZ}}$
mechanism for TOTIs (TOTSCs) }

\subsection{\label{subsec:Two-particle-Wess-Zumino-term-1}Two-particle Wess-Zumino term $\tilde{\mathrm{n}}_{\mathrm{WZ}}$ }

\begin{figure*}
\includegraphics[scale=1.1]{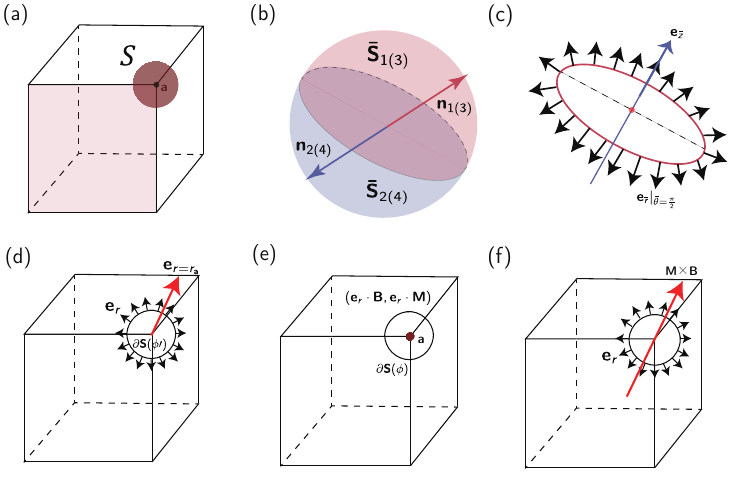}

\caption{\label{Fig2} Two-particle Wess-Zumino term $\tilde{\mathrm{n}}_{\mathrm{WZ}}$ mechanism.
(a) Area $\mathcal{\boldsymbol{S}}$ centered at corner $\mathrm{\boldsymbol{a}}$.
(b) Surface (pseudo)spin textures within $\mathcal{\boldsymbol{S}}$. The red and
bule hemispherical surfaces ($\mathcal{\boldsymbol{\bar{S}}}_{1(3)}$ and $\mathcal{\boldsymbol{\bar{S}}}_{2(4)}$)
denote the textures of (pseudo)spin polarizations $\mathrm{\boldsymbol{n}}_{1(3)}$
(red arrow) and $\mathrm{\boldsymbol{n}}_{2(4)}$(blue arrow) under the ideal condition,
respectively. Here $\mathcal{\boldsymbol{\bar{S}}}_{1(3)}+\mathcal{\boldsymbol{\bar{S}}}_{2(4)}$
make up a perfect unit sphere which renders the $\tilde{\mathrm{n}}_{\mathrm{WZ}}$.
(c) The $\tilde{\mathrm{n}}_{\mathrm{WZ}}$ can reduce to a winding number of $\boldsymbol{e_{\bar{r}}}|_{\bar{\theta}=\frac{\pi}{2}}$
on the equator of $\mathcal{\boldsymbol{\bar{S}}}_{1(3)}+\mathcal{\boldsymbol{\bar{S}}}_{2(4)}$.
Vector $\boldsymbol{e}_{r=r_{\mathrm{\boldsymbol{a}}}}$ in real space is taken as
$\boldsymbol{e}_{\bar{z}}$ in (pseudo)spin space. (d) The winding of $\boldsymbol{e_{\bar{r}}}|_{\bar{\theta}=\frac{\pi}{2}}$
can be mapped to that of $\boldsymbol{e}_{r}$ along the boundary $\partial\mathcal{\boldsymbol{S}}$
in real space. (e) A specialized mass field $(\mathrm{B}_{\phi},\mathrm{M}_{\phi})=(\boldsymbol{e}_{r}\cdot\mathbf{B},\boldsymbol{e}_{r}\cdot\boldsymbol{\mathrm{M}})$
is placed on the surface of 3D TI (TSC). (f) A nonzero CS term $\tilde{\theta}$
can be caused by the winding of $\boldsymbol{e}_{r}$ along $\partial\mathcal{\boldsymbol{S}}$.}
\end{figure*}

In this section, We derive the two-particle Wess-Zumino term $\text{ \ensuremath{\tilde{\mathrm{n}}_{\mathrm{WZ}}} }$
which is defined on the surface of 3D TIs (TSCs). As discussed in \citep{XJLiuPRB2022-1},
the trajectories of pseudospins $\boldsymbol{\mathrm{n}}_{\uparrow}$ and $\boldsymbol{\mathrm{n}}_{\downarrow}$
of 2D quantum spin Hall insualtos from one edge to adjacent one jointly form a closed
loop (circle) under the ideal condition, which enables the appearance of two-particel
Berry phase. Naturally, when we focus on 3D TIs (TSCs), the surface (pseudo)spin
textures around target corner are expected to constitute a perfect sphere and may
also contribute to an interesting topological invariant. To demonstrate this we concentrate
on an area $\mathcal{\boldsymbol{S}}$ of surface of 3D TIs (TSCs) with its center
locating in corner $\boldsymbol{a}$ {[}Fig.\ref{Fig2}(a){]} and study the topological
structure of surface (pseudo)spin textures defined on it. For a minimal 3D $\mathcal{T}$-invariant
3D TIs (TSCs), the spinor parts of topological surface states usually take the form
of 
\begin{equation}
|\tilde{\varepsilon}_{1}\rangle=\frac{1}{\sqrt{2}}\left(\begin{array}{c}
1\\
-\mathrm{i}
\end{array}\right)_{\sigma(\tau)}\otimes\left(\begin{array}{c}
\cos\theta/2\\
e^{i\varphi}\sin\theta/2
\end{array}\right)_{s};\quad|\tilde{\varepsilon}_{2}\rangle=\frac{1}{\sqrt{2}}\left(\begin{array}{c}
1\\
\mathrm{i}
\end{array}\right)_{\sigma(\tau)}\otimes\left(\begin{array}{c}
\sin\theta/2\\
-e^{i\varphi}\cos\theta/2
\end{array}\right)_{s},
\end{equation}
where $\theta(\varphi)$ represents the polar (azimuthal) angle in real space. Notice
that $\sigma$, $\tau$ and $s$ denote the pseudospin, Nambu and spin degrees of
freedom, respectively. Now we can define the surface (pseudo)spin polarizations:
$\boldsymbol{\mathrm{n}}_{1}=\langle\tilde{\varepsilon}_{1}|\boldsymbol{s}|\tilde{\varepsilon}_{1}\rangle$,
$\boldsymbol{\mathrm{n}}_{2}=\langle\tilde{\varepsilon}_{2}|\boldsymbol{s}|\tilde{\varepsilon}_{2}\rangle$.
Here $\boldsymbol{\mathrm{n}}_{1}=-\boldsymbol{\mathrm{n}}_{2}$ since $\langle\tilde{\varepsilon}_{1}|\boldsymbol{s}|\tilde{\varepsilon}_{1}\rangle=\langle\tilde{\varepsilon}_{2}|\mathcal{\hat{T}}^{-1}\boldsymbol{s}\mathcal{\hat{T}}|\tilde{\varepsilon}_{2}\rangle=-\langle\tilde{\varepsilon}_{2}|\boldsymbol{s}|\tilde{\varepsilon}_{2}\rangle$.
And we find that (pseudo)spin $\boldsymbol{\mathrm{n}}_{1}=(\sin\theta\cos\varphi,\sin\theta\sin\varphi,\cos\theta)$
is coincident with the surface normal vector $\boldsymbol{e}_{r}$ of 3D TIs (TSCs).
As shown in {[}Fig.\ref{Fig2}(b){]}, only one $\mathrm{\boldsymbol{n}}_{\boldsymbol{\nu}=(1,2)}$
within $\mathcal{\boldsymbol{S}}$ doesn't form a perfect sphere, but the joint textures
of two polarizations $\mathrm{\boldsymbol{n}}_{1}$ and $\mathrm{\boldsymbol{n}}_{2}$
can constitute a ``two-particle'' unit sphere $\boldsymbol{\bar{S}}_{1}+\boldsymbol{\bar{S}}_{2}$
in (pseudo)spin space under ideal condition, resembling a hedgehog ball. Thus we
can define a two-particle WZ term on $\mathcal{\boldsymbol{S}}$: 
\begin{align}
\mathrm{\tilde{n}}_{\mathrm{WZ}} & =\frac{1}{2\pi}\int_{\mathcal{\boldsymbol{S}}}\mathrm{d}\theta\mathrm{d}\varphi(\partial_{\theta}\tilde{\mathcal{A}}_{\varphi}-\partial_{\varphi}\tilde{\mathcal{A}}_{\theta}),\label{eq:14}
\end{align}
where $\tilde{\mathcal{A}}_{\theta}=\langle\tilde{\varepsilon}_{1}|\otimes\langle\tilde{\varepsilon}_{2}|(-\mathrm{i}\partial_{\theta})|\tilde{\varepsilon}_{1}\rangle\otimes|\tilde{\varepsilon}_{2}\rangle$,
$\tilde{\mathcal{A}}_{\varphi}=\langle\tilde{\varepsilon}_{1}|\otimes\langle\tilde{\varepsilon}_{2}|(-\mathrm{i}\partial_{\varphi})|\tilde{\varepsilon}_{1}\rangle\otimes|\tilde{\varepsilon}_{2}\rangle$
denote the two-particle Berry connections. Performing a transformation from real
space to (pseudo)spin space: 
\begin{align}
\langle\tilde{\varepsilon}_{1}|\otimes\langle\tilde{\varepsilon}_{2}|(-\mathrm{i}\partial_{\theta})|\tilde{\varepsilon}_{1}\rangle\otimes|\tilde{\varepsilon}_{2}\rangle & =\frac{\partial\bar{\theta}_{1}}{\partial\theta}\langle\tilde{\varepsilon}_{1}|(-\mathrm{i}\partial_{\bar{\theta}_{1}})|\tilde{\varepsilon}_{1}\rangle+\frac{\partial\bar{\theta}_{2}}{\partial\theta}\langle\tilde{\varepsilon}_{2}|(-\mathrm{i}\partial_{\bar{\theta}_{2}})|\tilde{\varepsilon}_{2}\rangle\nonumber \\
\langle\tilde{\varepsilon}_{1}|\otimes\langle\tilde{\varepsilon}_{2}|(-\mathrm{i}\partial_{\varphi})|\tilde{\varepsilon}_{1}\rangle\otimes|\tilde{\varepsilon}_{2}\rangle & =\frac{\partial\bar{\varphi}_{1}}{\partial\varphi}\langle\tilde{\varepsilon}_{1}|(-\mathrm{i}\partial_{\bar{\varphi}_{1}})|\tilde{\varepsilon}_{1}\rangle+\frac{\partial\bar{\varphi}_{2}}{\partial\varphi}\langle\tilde{\varepsilon}_{2}|(-\mathrm{i}\partial_{\bar{\varphi}_{2}})|\tilde{\varepsilon}_{2}\rangle,\label{eq:15}
\end{align}
we obtain $\tilde{\mathcal{A}}_{\theta}=\frac{\partial\bar{\theta}_{1}}{\partial\theta}\mathcal{\tilde{A}}_{\bar{\theta}_{1}}+\frac{\partial\bar{\theta}_{2}}{\partial\theta}\mathcal{\tilde{A}}_{\bar{\theta}_{2}}$,
$\tilde{\mathcal{A}}_{\varphi}=\frac{\partial\bar{\varphi}_{1}}{\partial\varphi}\mathcal{\tilde{A}}_{\bar{\varphi}_{1}}+\frac{\partial\bar{\varphi}_{2}}{\partial\varphi}\tilde{\mathcal{A}}_{\bar{\varphi}_{2}}$
with $\mathcal{\tilde{A}}_{\bar{\theta}_{i}(\bar{\varphi}_{i})}=\langle\tilde{\varepsilon}_{i}|(-\mathrm{i}\partial_{\bar{\theta}_{i}(\bar{\varphi}_{i})})|\tilde{\varepsilon}_{i}\rangle$.
Here we emphasize that $\bar{\theta}_{i}$ and $\bar{\varphi}_{i}$ are the polar
and azimuthal angles with respect to $\boldsymbol{e}_{\bar{z}}$ in (pseudo)spin
space. Notice that $\boldsymbol{e}_{\bar{z}}$ is chosen as the vector $\boldsymbol{e}_{r=r_{\mathrm{\boldsymbol{a}}}}$
from ``center of $\partial\mathcal{\boldsymbol{S}}$'' to corner $\mathrm{\boldsymbol{a}}$,
where ``center of $\partial\mathcal{\boldsymbol{S}}$'' is defined as the center
of section obtained by cutting corner $\mathrm{\boldsymbol{a}}$ along $\partial\mathcal{\boldsymbol{S}}$.
Substituting Eq.(\ref{eq:15}) into (\ref{eq:14}), we find that 
\begin{align}
\mathrm{\tilde{n}}_{\mathrm{WZ}}= & \frac{1}{2\pi}\int_{\mathcal{\boldsymbol{S}}}\mathrm{d}\theta\mathrm{d}\varphi[\partial_{\theta}\tilde{\mathcal{A}}_{\varphi}-\partial_{\varphi}\tilde{\mathcal{A}}_{\theta}]\nonumber \\
= & \frac{1}{2\pi}\int_{\mathcal{\boldsymbol{S}}}\mathrm{d}\theta\mathrm{d}\varphi[\partial_{\theta}(\frac{\partial\bar{\varphi}_{1}}{\partial\varphi}\mathcal{\tilde{A}}_{\bar{\varphi}_{1}}+\frac{\partial\bar{\varphi}_{2}}{\partial\varphi}\tilde{\mathcal{A}}_{\bar{\varphi}_{2}})-\partial_{\varphi}(\frac{\partial\bar{\theta}_{1}}{\partial\theta}\mathcal{\tilde{A}}_{\bar{\theta}_{1}}+\frac{\partial\bar{\theta}_{2}}{\partial\theta}\mathcal{\tilde{A}}_{\bar{\theta}_{2}})]\nonumber \\
= & \frac{1}{2\pi}\int_{\mathcal{\boldsymbol{S}}}\mathrm{d}\theta\mathrm{d}\varphi[(\partial_{\theta}\frac{\partial\bar{\varphi}_{1}}{\partial\varphi}\mathcal{\tilde{A}}_{\bar{\varphi}_{1}}-\partial_{\varphi}\frac{\partial\bar{\theta}_{1}}{\partial\theta}\mathcal{\tilde{A}}_{\bar{\theta}_{1}})+(\partial_{\theta}\frac{\partial\bar{\varphi}_{2}}{\partial\varphi}\tilde{\mathcal{A}}_{\bar{\varphi}_{2}}-\partial_{\varphi}\frac{\partial\bar{\theta}_{2}}{\partial\theta}\mathcal{\tilde{A}}_{\bar{\theta}_{2}})]\nonumber \\
= & \frac{1}{2\pi}\int_{\mathcal{\boldsymbol{S}}}\mathrm{d}\theta\mathrm{d}\varphi[(\frac{\partial\bar{\theta}_{1}}{\partial\theta}\frac{\partial\bar{\varphi}_{1}}{\partial\varphi}\partial_{\bar{\theta}_{1}}\mathcal{\tilde{A}}_{\bar{\varphi}_{1}}-\frac{\partial\bar{\theta}_{1}}{\partial\theta}\frac{\partial\bar{\varphi}_{1}}{\partial\varphi}\partial_{\bar{\varphi}_{1}}\mathcal{\tilde{A}}_{\bar{\theta}_{1}})+(\frac{\partial\bar{\theta}_{2}}{\partial\theta}\frac{\partial\bar{\varphi}_{2}}{\partial\varphi}\partial_{\bar{\theta}_{2}}\tilde{\mathcal{A}}_{\bar{\varphi}_{2}}-\frac{\partial\bar{\theta}_{2}}{\partial\theta}\frac{\partial\bar{\varphi}_{2}}{\partial\varphi}\partial_{\bar{\varphi}_{2}}\mathcal{\tilde{A}}_{\bar{\theta}_{2}})]\nonumber \\
= & \frac{1}{2\pi}\int_{\mathcal{\boldsymbol{S}}}\mathrm{d}\theta\mathrm{d}\varphi\frac{\partial\bar{\theta}_{1}}{\partial\theta}\frac{\partial\bar{\varphi}_{1}}{\partial\varphi}(\partial_{\bar{\theta}_{1}}\mathcal{\tilde{A}}_{\bar{\varphi}_{1}}-\partial_{\bar{\varphi}_{1}}\mathcal{\tilde{A}}_{\bar{\theta}_{1}})+\frac{1}{2\pi}\int_{\mathcal{\boldsymbol{S}}}\mathrm{d}\theta\mathrm{d}\varphi\frac{\partial\bar{\theta}_{2}}{\partial\theta}\frac{\partial\bar{\varphi}_{2}}{\partial\varphi}(\partial_{\bar{\theta}_{2}}\tilde{\mathcal{A}}_{\bar{\varphi}_{2}}-\partial_{\bar{\varphi}_{2}}\mathcal{\tilde{A}}_{\bar{\theta}_{2}})\nonumber \\
= & \frac{1}{2\pi}\int_{\mathcal{\boldsymbol{\bar{S}}}_{1}}\mathrm{d}\bar{\theta}_{1}\mathrm{d}\bar{\varphi}_{1}(\partial_{\bar{\theta}_{1}}\mathcal{\tilde{A}}_{\bar{\varphi}_{1}}-\partial_{\bar{\varphi}_{1}}\mathcal{\tilde{A}}_{\bar{\theta}_{1}})+\frac{1}{2\pi}\int_{\mathcal{\boldsymbol{\bar{S}}}_{2}}\mathrm{d}\bar{\theta}_{2}\mathrm{d}\bar{\varphi}_{2}(\partial_{\bar{\theta}_{2}}\tilde{\mathcal{A}}_{\bar{\varphi}_{2}}-\partial_{\bar{\varphi}_{2}}\mathcal{\tilde{A}}_{\bar{\theta}_{2}})\nonumber \\
= & \frac{1}{2\pi}\int_{\mathcal{\boldsymbol{\bar{S}}}_{1}}\mathrm{d}\bar{\theta}_{1}\mathrm{d}\bar{\varphi}_{1}\mathcal{\tilde{B}}_{\bar{\theta}_{1}\bar{\varphi}_{1}}+\frac{1}{2\pi}\int_{\mathcal{\boldsymbol{\bar{S}}}_{2}}\mathrm{d}\bar{\theta}_{2}\mathrm{d}\bar{\varphi}_{2}\tilde{\mathcal{B}}_{\bar{\theta}_{2}\bar{\varphi}_{2}}\nonumber \\
= & \frac{1}{2\pi}\int_{\mathcal{\boldsymbol{\bar{S}}}_{1}}\mathrm{d}\bar{\theta}_{1}\mathrm{d}\bar{\varphi}_{1}[-\mathrm{i}(\langle\partial_{\bar{\theta}_{1}}\tilde{\varepsilon}_{1}|\partial_{\bar{\varphi}_{1}}\tilde{\varepsilon}_{1}\rangle-\langle\partial_{\bar{\varphi}_{1}}\tilde{\varepsilon}_{1}|\partial_{\bar{\theta}_{1}}\tilde{\varepsilon}_{1}\rangle)]+\frac{1}{2\pi}\int_{\mathcal{\boldsymbol{\bar{S}}}_{2}}\mathrm{d}\bar{\theta}_{2}\mathrm{d}\bar{\varphi}_{2}[-\mathrm{i}(\langle\partial_{\bar{\theta}_{2}}\tilde{\varepsilon}_{2}|\partial_{\bar{\varphi}_{2}}\tilde{\varepsilon}_{2}\rangle-\langle\partial_{\bar{\varphi}_{2}}\tilde{\varepsilon}_{2}|\partial_{\bar{\theta}_{2}}\tilde{\varepsilon}_{2}\rangle)],
\end{align}
where $\bar{\theta}_{2}=\pi-\bar{\theta}_{1}$, $\bar{\varphi}_{2}=\pi+\bar{\varphi}_{1}$.
Using the relation $-\mathrm{i}(\langle\partial_{\theta_{\mu}}\varepsilon_{\mu}|\partial_{\varphi_{\mu}}\varepsilon_{\mu}\rangle-\langle\partial_{\varphi_{\mu}}\varepsilon_{\mu}|\partial_{\theta_{\mu}}\varepsilon_{\mu}\rangle)=(-\mathrm{i})\mathrm{tr}\epsilon_{\theta_{\mu}\varphi_{\mu}}(P_{\mu}\partial_{\theta_{\mu}}P_{\mu}\partial_{\varphi_{\mu}}P_{\mu})=\frac{\mu}{2}\boldsymbol{\boldsymbol{e}_{r}}\cdot(\partial_{\theta_{\mu}}\boldsymbol{\boldsymbol{e}_{r}}\times\partial_{\varphi_{\mu}}\boldsymbol{\boldsymbol{e}_{r}})$
with $\mu=1,2$ and $P_{\mu}=\frac{1}{2}(1+\mu\sum_{i}e_{i}s_{i})$, the two-particle
Wess-Zumino term $\mathrm{\tilde{n}}_{\mathrm{WZ}}$ becomes 
\begin{align}
\mathrm{\tilde{n}}_{\mathrm{WZ}} & =\frac{1}{4\pi}\{\int_{\mathcal{\boldsymbol{\bar{S}}}_{1}}\mathrm{d}\bar{\theta}_{1}\mathrm{d}\bar{\varphi}_{1}[\mathrm{\boldsymbol{n}}_{1}\cdot(\partial_{\bar{\theta}_{1}}\mathrm{\boldsymbol{n}}_{1}\times\partial_{\bar{\varphi}_{1}}\mathrm{\boldsymbol{n}}_{1})]+\int_{\mathcal{\boldsymbol{\bar{S}}}_{2}}\mathrm{d}\bar{\theta}_{2}\mathrm{d}\bar{\varphi}_{2}[\mathrm{\boldsymbol{n}}_{2}\cdot(\partial_{\bar{\theta}_{2}}\mathrm{\boldsymbol{n}}_{2}\times\partial_{\bar{\varphi}_{2}}\mathrm{\boldsymbol{n}}_{2})]\}.\label{eq:17}
\end{align}
Here $\mathcal{\boldsymbol{\bar{S}}}_{1}+\mathcal{\boldsymbol{\bar{S}}}_{2}$ represents
a unit spherical surface in (pseudo)spin space with $|\mathcal{\boldsymbol{\bar{S}}}_{1}|=|\mathcal{\boldsymbol{\bar{S}}}_{2}|$.
Since $\int_{\mathcal{\bar{S}}_{1}}\mathrm{d}\bar{\Omega}_{1}+\int_{\mathcal{\bar{S}}_{2}}\mathrm{d}\bar{\Omega}_{2}=\int_{\mathcal{\bar{S}}_{1}}\mathrm{d}\bar{\Omega}_{1}+\int_{-\mathcal{\bar{S}}_{2}}(-\mathrm{d}\bar{\Omega}_{1})$,
we have 
\begin{align}
\mathrm{\tilde{n}}_{\mathrm{WZ}} & =\frac{1}{4\pi}\int_{\mathcal{\boldsymbol{\bar{S}}}_{1}}\mathrm{d}\bar{\theta}_{1}\mathrm{d}\bar{\varphi}_{1}[\mathrm{\boldsymbol{n}}_{1}\cdot(\partial_{\bar{\theta}_{1}}\mathrm{\boldsymbol{n}}_{1}\times\partial_{\bar{\varphi}_{1}}\mathrm{\boldsymbol{n}}_{1})]+\frac{1}{4\pi}\int_{-\mathcal{\boldsymbol{\bar{S}}}_{2}}\mathrm{d}\bar{\theta}_{1}\frac{\partial\bar{\theta}_{2}}{\partial\bar{\theta}_{1}}\mathrm{d}\bar{\varphi}_{1}\frac{\partial\bar{\varphi}_{2}}{\partial\bar{\varphi}_{1}}[\mathrm{\boldsymbol{n}}_{2}\cdot(\frac{\partial\bar{\theta}_{2}}{\partial\bar{\theta}_{1}}\partial_{\bar{\theta}_{1}}\mathrm{\boldsymbol{n}}_{2}\times\frac{\partial\bar{\varphi}_{2}}{\partial\bar{\varphi}_{1}}\partial_{\bar{\varphi}_{1}}\mathrm{\boldsymbol{n}}_{2})]\nonumber \\
 & =\frac{1}{4\pi}\int_{\mathcal{\boldsymbol{\bar{S}}}_{1}}\mathrm{d}\bar{\theta}_{1}\mathrm{d}\bar{\varphi}_{1}[\mathrm{\boldsymbol{n}}_{1}\cdot(\partial_{\bar{\theta}_{1}}\mathrm{\boldsymbol{n}}_{1}\times\partial_{\bar{\varphi}_{1}}\mathrm{\boldsymbol{n}}_{1})]-\frac{1}{4\pi}\int_{-\mathcal{\boldsymbol{\bar{S}}}_{2}}\mathrm{d}\bar{\theta}_{1}\mathrm{d}\bar{\varphi}_{1}[\mathrm{\boldsymbol{n}}_{1}\cdot(\partial_{\bar{\theta}_{1}}\mathrm{\boldsymbol{n}}_{1}\times\partial_{\bar{\varphi}_{1}}\mathrm{\boldsymbol{n}}_{1})]\nonumber \\
 & =\frac{1}{4\pi}\int_{\mathcal{\boldsymbol{\bar{S}}}_{1}-(-\mathcal{\boldsymbol{\bar{S}}}_{2})}\mathrm{d}\bar{\theta}_{1}\mathrm{d}\bar{\varphi}_{1}[\mathrm{\boldsymbol{n}}_{1}\cdot(\partial_{\bar{\theta}_{1}}\mathrm{\boldsymbol{n}}_{1}\times\partial_{\bar{\varphi}_{1}}\mathrm{\boldsymbol{n}}_{1})]\nonumber \\
 & =\frac{1}{4\pi}\int_{\mathrm{sphere}}\mathrm{d}\bar{\theta}\mathrm{d}\bar{\varphi}[\boldsymbol{e_{\bar{r}}}\cdot(\partial_{\bar{\theta}}\boldsymbol{e_{\bar{r}}}\times\partial_{\bar{\varphi}}\boldsymbol{e_{\bar{r}}})].\label{eq:18}
\end{align}
As we can see, Eq.(\ref{eq:18}) returns to the general form of Wess-Zumino term
$\mathrm{n}_{\mathrm{WZ}}$, which explains that why Eq.(\ref{eq:17}) can be dubbed
the two-particle Wess-Zumino term $\mathrm{\tilde{n}}_{\mathrm{WZ}}$. 

\begin{figure*}
\includegraphics[scale=1.1]{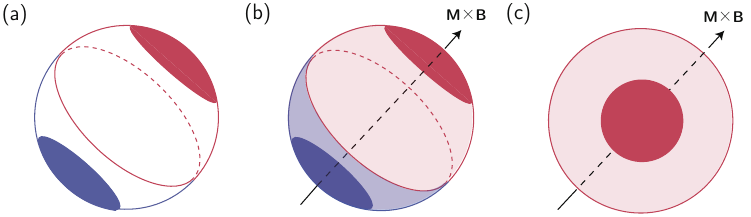}

\caption{\label{Fig3} (a) Schematic of surface (pseudo)spin textures for non-closed case.
Here the dark red and dark blue regions denote the textures of (pseudo)spins $\mathrm{\boldsymbol{n}}_{1(3)}$
and $\mathrm{\boldsymbol{n}}_{2(4)}$, respectively. (b) The situation where $\tilde{\mathrm{n}}_{\mathrm{WZ}}$
can be well-defined. Since the real surface (pseudo)spin texture regions are non-closed,
we need to add the auxiliary textures (light red and light blue regions) to complement
real regions and achieve a perfect unit sphere, which is dubbed ``continuation''.
And as long as the mass field vortex exists in real regions (the vector $\boldsymbol{e}_{\boldsymbol{\mathrm{M}}\times\boldsymbol{\mathrm{B}}}$
threads the dark color regions), we can obtain a nonzero two-particle WZ terms. (c)
The situation where we can't get a well-defined $\tilde{\mathrm{n}}_{\mathrm{WZ}}$
since the mass field vortex locates in auxiliary regions.}
\end{figure*}

\subsection{\label{subsec:Two-particle-Wess-Zumino-term}Two-particle Wess-Zumino term $\tilde{\mathrm{n}}_{\mathrm{WZ}}$
mechanism}

Before proceeding, we emphasize that 3D TIs (TSCs) described by a minimal $4\times4$
tight-binding Hamiltonain $\mathcal{H}(\boldsymbol{k})$ can't harbour the 3rd-order
zero corner modes as only one effective surface mass term $\mathrm{m}\sigma_{k}$
can gap out the surface Dirac (Majorana) cone captured by $\mathcal{H}=k_{x}\sigma_{i}+k_{y}\sigma_{j}$,
where $\sigma_{i}$ is the Pauli Matrix. In order to construct the TOTIs and TOTSCs,
we need to stack two copies of minimal 3D TIs (TSCs), which is dubbed the ``double-TI
(TSC)'' construction \citep{KhalafPRX2018-1,KhalafPRB2018-1,BernevigPRB2019-1}.
And the surface Dirac (Majorana) cone is given by $\mathcal{H}^{\prime}=k_{x}\Gamma_{1}+k_{y}\Gamma_{2}$
with $\boldsymbol{\Gamma}$ been gamma matrices. Then we can apply a mass field $\mathrm{m}_{\mathrm{B}}\Gamma_{3}+\mathrm{m}_{\mathrm{M}}\Gamma_{4}$
to gap out such surface Dirac (Majorana) cone, leading to the zero energy corner
modes. 

Now we show the $\tilde{\mathrm{n}}_{\mathrm{WZ}}$ mechanism for 3rd-order zero
corner modes of TOTIs (TOTSCs). First, we construct the 3D total TIs (TSCs) by stacking
two identical time-reversal invariant TIs (TSCs) $\mathcal{H}_{\mathrm{TI(TSC)}}^{\mathrm{total}}=\mathcal{H}_{\mathrm{TI(TSC)}}^{\mathrm{\alpha}}\oplus\mathcal{H}_{\mathrm{TI(TSC)}}^{\mathrm{\beta}}$,
where $\mathcal{H}_{\mathrm{TI(TSC)}}^{\mathrm{\alpha}(\beta)}$ is the minimal $4\times4$
tight-binding Hamiltonain. Hence the surface (pseudo)spin polarizations are grouped
into $\{\boldsymbol{\mathrm{n}}_{1},\boldsymbol{\mathrm{n}}_{2}\}_{\alpha}$ and
$\{\boldsymbol{\mathrm{n}}_{3},\boldsymbol{\mathrm{n}}_{4}\}_{\beta}$ with $\boldsymbol{\mathrm{n}}_{1}=-\boldsymbol{\mathrm{n}}_{2}=\boldsymbol{\mathrm{n}}_{3}=-\boldsymbol{\mathrm{n}}_{4}=\boldsymbol{e}_{r}$.
Correspondingly, the surface (pseudo)spin textures within $\mathcal{\boldsymbol{S}}$
render two coincident unit spheres under the ideal condition. Thus we have
\begin{align}
\mathrm{\tilde{n}}_{\mathrm{WZ}}^{(\alpha)}=\mathrm{\tilde{n}}_{\mathrm{WZ}}^{(\beta)}= & \frac{1}{4\pi}\{\int_{\mathcal{\boldsymbol{\bar{S}}}_{1}}\mathrm{d}\bar{\theta}_{1}\mathrm{d}\bar{\varphi}_{1}[\mathrm{\boldsymbol{n}}_{1}\cdot(\partial_{\bar{\theta}_{1}}\mathrm{\boldsymbol{n}}_{1}\times\partial_{\bar{\varphi}_{1}}\mathrm{\boldsymbol{n}}_{1})]+\int_{\mathcal{\boldsymbol{\bar{S}}}_{2}}\mathrm{d}\bar{\theta}_{2}\mathrm{d}\bar{\varphi}_{2}[\mathrm{\boldsymbol{n}}_{2}\cdot(\partial_{\bar{\theta}_{2}}\mathrm{\boldsymbol{n}}_{2}\times\partial_{\bar{\varphi}_{2}}\mathrm{\boldsymbol{n}}_{2})]\nonumber \\
= & \frac{1}{4\pi}\{\int_{\mathcal{\boldsymbol{\bar{S}}}_{3}}\mathrm{d}\bar{\theta}_{3}\mathrm{d}\bar{\varphi}_{3}[\mathrm{\boldsymbol{n}}_{3}\cdot(\partial_{\bar{\theta}_{3}}\mathrm{\boldsymbol{n}}_{3}\times\partial_{\bar{\varphi}_{3}}\mathrm{\boldsymbol{n}}_{3})]+\int_{\mathcal{\boldsymbol{\bar{S}}}_{4}}\mathrm{d}\bar{\theta}_{4}\mathrm{d}\bar{\varphi}_{4}[\mathrm{\boldsymbol{n}}_{4}\cdot(\partial_{\bar{\theta}_{4}}\mathrm{\boldsymbol{n}}_{4}\times\partial_{\bar{\varphi}_{4}}\mathrm{\boldsymbol{n}}_{4})]\nonumber \\
= & \frac{1}{4\pi}\int_{\mathrm{sphere}}\mathrm{d}\bar{\theta}\mathrm{d}\bar{\varphi}[\boldsymbol{e_{\bar{r}}}\cdot(\partial_{\bar{\theta}}\boldsymbol{e_{\bar{r}}}\times\partial_{\bar{\varphi}}\boldsymbol{e_{\bar{r}}})],\label{eq:19}
\end{align}
where $\bar{\theta}_{1}=\bar{\theta}_{3}$, $\bar{\theta}_{2}=\bar{\theta}_{4}$,
$\bar{\varphi}_{1}=\bar{\varphi}_{3}$, $\bar{\varphi}_{2}=\bar{\varphi}_{4}$. For
simplicity, we only take the $\mathrm{\tilde{n}}_{\mathrm{WZ}}^{(\alpha)}$ as example
to proceed the derivation, but the results are also the same with $\mathrm{\tilde{n}}_{\mathrm{WZ}}^{(\beta)}$.
As shown in Fig.\ref{Fig2}(c), Eq.(\ref{eq:19}) can reduce to a line integral 
\begin{align}
\mathrm{\tilde{n}}_{\mathrm{WZ}}^{(\alpha)}=\mathrm{\tilde{n}}_{\mathrm{WZ}}^{(\beta)}= & \frac{1}{2\pi}\oint_{\mathrm{equator}}\mathrm{d}\bar{\varphi}[\boldsymbol{e_{\bar{z}}}\cdot(\boldsymbol{e_{\bar{r}}}|_{\bar{\theta}=\frac{\pi}{2}}\times\partial_{\bar{\varphi}}\boldsymbol{e_{\bar{r}}}|_{\bar{\theta}=\frac{\pi}{2}})],\label{eq:20}
\end{align}
which indicates that the nonzero two-particle Wess-Zumino term ($\mathrm{\tilde{n}}_{\mathrm{WZ}}^{(\alpha)}=\mathrm{\tilde{n}}_{\mathrm{WZ}}^{(\beta)}=1$)
can contribute to a nonzero winding numbers of surface normal vector $\boldsymbol{e_{\bar{r}}}|_{\bar{\theta}=\frac{\pi}{2}}$
along the equator of unit sphere $\bar{\mathcal{\boldsymbol{S}}}_{1(3)}+\bar{\mathcal{\boldsymbol{S}}}_{2(4)}$.
Transforming such integral from (pseudo)spin space to real space, the vector $\boldsymbol{e_{\bar{r}}}|_{\bar{\theta}=\frac{\pi}{2}}$
is correspondingly mapped to vector $\boldsymbol{e}_{r}$ on the boundary of $\mathcal{\boldsymbol{S}}$
{[}Fig.\ref{Fig2}(d){]}: 
\begin{align}
\mathrm{\tilde{n}}_{\mathrm{WZ}}^{(\alpha)}=\mathrm{\tilde{n}}_{\mathrm{WZ}}^{(\beta)}= & \frac{1}{2\pi}\oint_{\mathrm{equator}}\mathrm{d}\bar{\varphi}[\boldsymbol{e_{\bar{z}}}\cdot(\boldsymbol{e_{\bar{r}}}|_{\bar{\theta}=\frac{\pi}{2}}\times\partial_{\bar{\varphi}}\boldsymbol{e_{\bar{r}}}|_{\bar{\theta}=\frac{\pi}{2}})]=\frac{1}{2\pi}\oint_{\partial\boldsymbol{S}(\phi^{\prime})}\mathrm{d}\phi^{\prime}[\boldsymbol{e}_{r=r_{\mathrm{\boldsymbol{a}}}}\cdot(\boldsymbol{e}_{r}\times\frac{\partial\bar{\varphi}}{\partial\phi^{\prime}}\partial_{\bar{\varphi}}\boldsymbol{e}_{r})]\nonumber \\
= & \frac{1}{2\pi}\oint_{\partial\boldsymbol{S}(\phi^{\prime})}\mathrm{d}\phi^{\prime}[\boldsymbol{e}_{r=r_{\mathrm{\boldsymbol{a}}}}\cdot(\boldsymbol{e}_{r}\times\partial_{\phi^{\prime}}\boldsymbol{e}_{r})].\label{eq:21}
\end{align}
Next, we discuss the connection between two-particle WZ term $\tilde{\mathrm{n}}_{\mathrm{WZ}}$
and CS term $\tilde{\theta}$. According to (\ref{sec:Chern-Simons-term-}), we can
gap out the surface Dirac (Majorana) cone of total TIs (TSCs) $\mathcal{H}_{\mathrm{TI(TSC)}}^{\mathrm{total}}$
and realize the TOTIs (TOTSCs) by applying a non-trivial surface effective mass field
$\mathrm{\boldsymbol{m}}=(\mathrm{B}_{\phi},\mathrm{M}_{\phi})$, and the surface
effective Hamiltonian is usually given by $\mathcal{H}_{\mathrm{eff}}=-tk_{\varphi}\Gamma_{1}+tk_{\theta}\Gamma_{2}+\mathrm{B}_{\phi}\Gamma_{3}+\mathrm{M}_{\phi}\Gamma_{4}$.
As a result, the Chern-Simons term $\tilde{\theta}$ with respect to target corner
$\mathrm{\boldsymbol{a}}$ is computed as $\tilde{\theta}=\pi[\frac{1}{2\pi}\oint_{\partial\boldsymbol{S}(\phi)}\mathrm{d}\phi\frac{\mathrm{M}_{\phi}\partial_{\phi}(\mathrm{B}_{\phi})-\mathrm{B}_{\phi}\partial_{\phi}(\mathrm{M}_{\phi})}{\mathrm{B}_{\phi}^{2}+\mathrm{M}_{\phi}^{2}}]$.
Here the Dirac (Majorana) cone winding number has been taken as $\mathcal{W}_{\mathrm{\boldsymbol{h}}}=1$,
and $\partial\mathcal{\boldsymbol{S}}$ represents the boundary of area $\mathcal{\boldsymbol{S}}$.
As sketched in Fig.\ref{Fig2}(e) and (f), we consider a kind of specialized surface
mass field $(\mathrm{B}_{\phi},\mathrm{M}_{\phi})=(\boldsymbol{e}_{r}\cdot\mathbf{B},\boldsymbol{e}_{r}\cdot\boldsymbol{\mathrm{M}})$
with $\boldsymbol{\mathrm{B}}$ and $\boldsymbol{\mathrm{M}}$ been constant unit
vectors, thus the Chern-Simons term $\tilde{\theta}$ becomes 
\begin{align}
\tilde{\theta}= & \pi\{\frac{1}{2\pi}\oint_{\partial\boldsymbol{S}(\phi)}\mathrm{d}\phi[\frac{(\boldsymbol{e}_{r}\cdot\boldsymbol{\mathrm{M}})(\partial_{\phi}\boldsymbol{e}_{r}\cdot\mathbf{B})-(\boldsymbol{e}_{r}\cdot\mathbf{B})(\partial_{\phi}\boldsymbol{e}_{r}\cdot\boldsymbol{\mathrm{M}})}{(\boldsymbol{e}_{r}\cdot\mathbf{B})^{2}+(\boldsymbol{e}_{r}\cdot\boldsymbol{\mathrm{M}})^{2}}]\}=\frac{1}{2}\oint_{\partial\boldsymbol{S}(\phi)}\mathrm{d}\phi[\boldsymbol{e}_{\boldsymbol{\mathrm{M}}\times\boldsymbol{\mathrm{B}}}\cdot(\boldsymbol{\boldsymbol{e}_{r}}\times\partial_{\phi}\boldsymbol{\boldsymbol{e}_{r}})]\frac{|\boldsymbol{\mathrm{M}}\times\boldsymbol{\mathrm{B}}|}{\mathrm{B}_{\phi}^{2}+\mathrm{M}_{\phi}^{2}}\label{eq:22}
\end{align}
As we can see, the two-particle WZ term $\mathrm{\tilde{n}}_{\mathrm{WZ}}$ in Eq.(\ref{eq:21})
is closely related to Chern-Simons term $\tilde{\theta}$ in Eq.(\ref{eq:22}). When
$|\boldsymbol{e}_{r=r_{\mathrm{\boldsymbol{a}}}}\cdot\boldsymbol{e}_{\boldsymbol{\mathrm{M}}\times\boldsymbol{\mathrm{B}}}|=1$
and $\boldsymbol{\mathrm{M}}\perp\boldsymbol{\mathrm{B}}$, we have $|\tilde{\theta}|=\pi|\mathrm{\tilde{n}}_{\mathrm{WZ}}^{(\alpha)}|=\pi|\mathrm{\tilde{n}}_{\mathrm{WZ}}^{(\beta)}|$.
Notice that this equality is also applied to a more general case: $|\boldsymbol{e}_{r=r_{\mathrm{\boldsymbol{a}}}}\cdot\boldsymbol{e}_{\boldsymbol{\mathrm{M}}\times\boldsymbol{\mathrm{B}}}|\leq1$
and $\boldsymbol{\mathrm{M}}\nparallel\boldsymbol{\mathrm{B}}$. This result reflects
that the two-particle WZ terms $\mathrm{\tilde{n}}_{\mathrm{WZ}}$ can render a nonzero
Chern-Simons term $\tilde{\theta}$ under the specialized effective mass field $(\boldsymbol{e}_{r}\cdot\mathbf{B},\boldsymbol{e}_{r}\cdot\boldsymbol{\mathrm{M}})$.
Namely, we can predict the existence of zero corner modes of TOTIs (TOTSCs) in terms
of $\mathrm{\tilde{n}}_{\mathrm{WZ}}$ which is derived from the surface (pseudo)spin
textures around target corner. 

However, as shown in {[}Fig.\ref{Fig3}(a){]}, the surface (pseudo)spin textues defined
on $\mathcal{\boldsymbol{S}}$ are usually not closed (dark red and dark blue spherical
crowns). Are the conclusions above still valid? The answer is yes. To demonstrate
this we will resort to the method: ``continuation'', which means the auxiliary
(pseudo)spin textures {[}light red and light blue regions in Fig.\ref{Fig3}(b){]}
are added to complement the real (pseudo)spin textures, rendering a perfect unit
sphere. Remarkably, the surface effective mass nodes must be guaranteed to locate
in real (pseudo)spin texture regions. With this method, the two-particle WZ terms
$\tilde{\mathrm{n}}_{\mathrm{WZ}}$ can be well-defined. For example, as plotted
in {[}Fig.\ref{Fig3}(b){]}, $\boldsymbol{e}_{\boldsymbol{\mathrm{M}}\times\boldsymbol{\mathrm{B}}}$
pierces the dark red and dark blue actual (pseudo)spin texture regions, making the
effective mass field vortex exist in them. Thus we have $\tilde{\mathrm{n}}_{\mathrm{WZ}}^{\alpha}=\tilde{\mathrm{n}}_{\mathrm{WZ}}^{\beta}=1$
after ``continuation''. In contrast, the two-particle WZ term can't be well-defined
and disappear for the case shown in {[}Fig.\ref{Fig3}(c){]}. In conclusion, the
two-particle WZ $\tilde{\mathrm{n}}_{\mathrm{WZ}}$ term mechanism can be stated
as follow: if the non-trivial (pseudo)spin textures around target corner support
the nonzero $\tilde{\mathrm{n}}_{\mathrm{WZ}}$, a zero energy bound state localizes
at the corner with specialized surface mass field $(\boldsymbol{e}_{r}\cdot\mathbf{B},\boldsymbol{e}_{r}\cdot\boldsymbol{\mathrm{M}})$
applied. 

\section{\label{sec:Feasible physical realizations for TOTIs (TOTSCs)}Feasible physical
realizations for TOTIs (TOTSCs)}

The surface Chern-Simons theory ($\tilde{\theta}$ characterization and $\tilde{\mathrm{n}}_{\mathrm{WZ}}$
mechanism) provides a generic and intuitive 2D effective surface theoretical description
for the emergence of zero energy corner modes of TOTIs (TOTSCs), which opens a new
avenue toward the realizations of TOTIs and TOTSCs and guides us to search for the
underlying material candidates. In the following, we provide the feasible physical
realization schemes respectively for TOTIs and TOTSCs. 

\begin{figure*}
\includegraphics{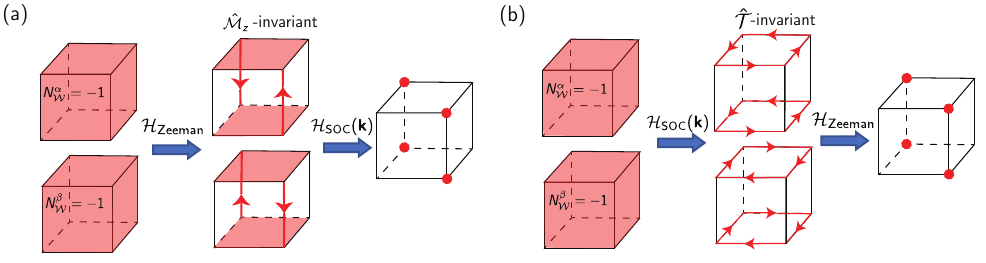}

\caption{\label{Fig4} Symmetry analysis for Majorana zero corner modes of TOTSCs. (a) The
total TSC is constructed by stacking two copies of minimal $\mathcal{T}$-invariant
TSCs. Applying an in-plane Zeeman field produces the helical hinge modes protected
by $\mathcal{M}_{z}$ symmetry. Then SOC term opens gap and introduces mass-inversion
bewteen adjacent hinges, leading to Majorana zero corner modes. (b) Alternatively,
adding SOC term breaks $\mathcal{M}_{z}$ symmetry while keeps $\mathcal{T}$ symmetry,
thus a pair of helical hinge modes protected by $\mathcal{T}$ appear. Then an in-plane
Zeeman term lifts $\mathcal{T}$ symmetry and gives rise to Majorana zero corner
modes.}
\end{figure*}

\subsection{\label{subsec:Realization scheme for TOTSCsfor}Realization scheme for TOTSCs }

\subsubsection{$\tilde{\theta}$ characterization}

Let us start from the Hamiltonian for TOTSCs: 
\begin{align}
\mathcal{H}_{\mathrm{TOTSC}}(\boldsymbol{k}) & =\mathcal{H}_{\mathrm{TSC}}^{\mathrm{total}}(\boldsymbol{k})+\mathcal{H}_{\mathrm{Zeeman}}+\mathcal{H}_{\mathrm{SOC}}(\boldsymbol{k}).\label{eq:23}
\end{align}
The 3D total TSC, which is constructed by stacking two copies of minimal 3D TSCs:
$\mathcal{H}_{\mathrm{TSC}}^{\mathrm{total}}(\boldsymbol{k})=\mathcal{H}_{\mathrm{TSC}}^{\alpha}(\boldsymbol{k})\oplus\mathcal{H}_{\mathrm{TSC}}^{\beta}(\boldsymbol{k})$,
is described by Hamiltonian: $\mathcal{H}_{\mathrm{TSC}}^{\mathrm{total}}(\boldsymbol{k})=\frac{\Delta_{p}}{k_{F}}\sum_{j=x,y}\sin k_{j}\sigma_{0}\tau_{x}s_{j}+\frac{\Delta_{p}}{k_{F}}\sin k_{z}\sigma_{z}\tau_{x}s_{z}-(\mu-\frac{3}{m_{\ast}}+\frac{1}{m_{\ast}}\sum_{j=x,y,z}\cos k_{j})\sigma_{z}\tau_{z}s_{0}$,
where $\{\boldsymbol{\sigma}\}$, $\{\boldsymbol{\tau}\}$ and $\{\boldsymbol{s}\}$
represent the Pauli matrices acting on orbital, Nambu and spin space, respectively.
Furthermore, the Zeeman term reads $\mathcal{H}_{\mathrm{Zeeman}}=\boldsymbol{B}_{\parallel}\cdot\boldsymbol{s}$
with $\boldsymbol{B}_{\parallel}=(\frac{\sqrt{2}}{2},\frac{\sqrt{2}}{2},0)$ and
the spin-orbit coupling (SOC) term $\mathcal{H}_{\mathrm{SOC}}(\boldsymbol{k})$
keeps the time-reversal symmetry $\hat{\mathcal{T}}=-\mathrm{i}s_{y}\kappa$ since
$\hat{\mathcal{T}}^{-1}\mathcal{H}_{\mathrm{SOC}}(\boldsymbol{k})\hat{\mathcal{T}}=\hat{\mathcal{T}}^{-1}[-\lambda_{so,0}+\lambda_{so,1}(\cos k_{x}+\cos k_{y})]\sigma_{y}s_{z}\hat{\mathcal{T}}=\mathcal{H}_{\mathrm{SOC}}(-\boldsymbol{k})$.
Now we use $\tilde{\theta}$ characterization to identify the 3rd-order zero energy
corner modes. We solve the topological surface states from $\mathcal{H}_{\mathrm{TSC}}^{\mathrm{total}}(\boldsymbol{k})$
for surface $\boldsymbol{\Sigma}(\theta=\frac{\pi}{2},\phi)$ in cylindrical coordinate
system: 
\begin{align}
|\epsilon_{1}\rangle=\frac{1}{2}\left(\begin{array}{c}
1\\
0
\end{array}\right)_{\sigma}\otimes\left(\begin{array}{c}
1\\
\mathrm{i}
\end{array}\right)_{\tau}\otimes\left(\begin{array}{c}
1\\
+e^{\mathrm{i}\phi}
\end{array}\right)_{s};\quad & |\epsilon_{2}\rangle=\frac{1}{2}\left(\begin{array}{c}
1\\
0
\end{array}\right)_{\sigma}\otimes\left(\begin{array}{c}
1\\
\mathrm{-i}
\end{array}\right)_{\tau}\otimes\left(\begin{array}{c}
1\\
-e^{\mathrm{i}\phi}
\end{array}\right)_{s}\nonumber \\
|\epsilon_{3}\rangle=\frac{1}{2}\left(\begin{array}{c}
0\\
1
\end{array}\right)_{\sigma}\otimes\left(\begin{array}{c}
1\\
\mathrm{i}
\end{array}\right)_{\tau}\otimes\left(\begin{array}{c}
1\\
-e^{\mathrm{i}\phi}
\end{array}\right)_{s};\quad & |\epsilon_{4}\rangle=\frac{1}{2}\left(\begin{array}{c}
0\\
1
\end{array}\right)_{\sigma}\otimes\left(\begin{array}{c}
1\\
-\mathrm{i}
\end{array}\right)_{\tau}\otimes\left(\begin{array}{c}
1\\
+e^{\mathrm{i}\phi}
\end{array}\right)_{s},\label{eq:24}
\end{align}
where $\phi$ is the azimuthal angle. And we can readily obtain the effective surface
Hamiltonian $\mathcal{H}_{\mathrm{effective}}=t_{p}k_{\phi}\tilde{\sigma}_{z}\tilde{\tau}_{x}+t_{p}k_{z}\tilde{\sigma}_{z}\tilde{\tau}_{y}+\mathrm{m}_{\phi}\tilde{\sigma}_{z}\tilde{\tau}_{z}+\mathrm{m}_{\theta}\tilde{\sigma}_{y}\tilde{\tau}_{0}$
with effective velocity $t_{p}=\frac{\Delta_{p}}{k_{F}}$. Since magnetic field $\boldsymbol{B}_{\parallel}=(\frac{\sqrt{2}}{2},\frac{\sqrt{2}}{2},0)$
is deposited on the $x-y$ plane, the resulting effective mass $\mathrm{m}_{\phi}=\boldsymbol{B}_{\parallel}\cdot\boldsymbol{e_{\rho}}$
forms a domain wall spreading along the meridian $\mathrm{\boldsymbol{l}}_{\phi=3\pi/4}$.
Moreover, the effective mass $\mathrm{m}_{\theta}$ generated by $\mathcal{H}_{\mathrm{SOC}}(\boldsymbol{k})$
is dependent on $\theta$ ($\mathrm{m}_{\theta}=-\lambda_{so,0}+2\lambda_{so,1}-\lambda_{so,1}\frac{\frac{4\mu}{m_{\ast}}-\frac{m_{\ast}^{2}}{4}-1}{2}\sin^{2}\theta$)
and its mass domain walls are present at two parallels $\mathrm{\boldsymbol{l}}_{\theta_{1}}$
and $\mathrm{\boldsymbol{l}}_{\theta_{2}}$. Then we can get four intersections of
$\mathrm{\boldsymbol{l}}_{\phi=3\pi/4}$ and $\mathrm{\boldsymbol{l}}_{\theta_{1}},\mathrm{\boldsymbol{l}}_{\theta_{2}}$,
which serve as the vortices of surface mass field $(\mathrm{m}_{\phi},\mathrm{m}_{\theta})$
here and contribute to the nonzero mass field winding numbers. Notice that the surface
Majorana cone winding number has been taken as $\mathcal{W}_{\mathrm{\boldsymbol{h}}}=1$.
As a result, there are totally four zero energy bound states at the intersections
according to $\tilde{\theta}$ characterization, which means the TOTSCs dictated
by Eq.(\ref{eq:23}) features four Majorana zero corner modes. The numerical results
displayed in Fig.3(d) of main text confirm this prediction. 

\subsubsection{Symmetry analysis}

Finally, we show that the existence of Majorana zero corner modes can also be deduced
from symmetry analysis. It's known that the 3D total TSC preserves time-reversal
symmetry as $\hat{\mathcal{T}}^{-1}\mathcal{H}_{\mathrm{TSC}}^{\mathrm{total}}(\boldsymbol{k})\hat{\mathcal{T}}=\mathcal{H}_{\mathrm{TSC}}^{\mathrm{total}}(-\boldsymbol{k})$.
Adding in-plane magnetic field $\boldsymbol{B}_{\parallel}$, $\hat{\mathcal{T}}$
symmetry is broken and a pair of helical hinge states which result from the sign-inversion
of $\mathrm{B}_{\phi}=\boldsymbol{B}\cdot\boldsymbol{e_{\rho}}$ propagate along
$z$ direction. Notice that mirror symmetry $\hat{\mathcal{M}}_{z}=\sigma_{x}\tau_{x}$
can relate helical hinge modes since BdG Hamiltonian $\mathcal{H}_{\mathrm{BdG}}(\boldsymbol{k})=\mathcal{H}_{\mathrm{TSC}}^{\mathrm{total}}(\boldsymbol{k})+\mathcal{H}_{\mathrm{Zeeman}}$
is $\hat{\mathcal{M}}_{z}$ invariant: $\hat{\mathcal{M}}_{z}^{-1}\mathcal{H}_{\mathrm{BdG}}(k_{x},k_{y},k_{z})\hat{\mathcal{M}}_{z}=\mathcal{H}_{\mathrm{BdG}}(k_{x},k_{y},-k_{z})$.
In other words, these gapless hinge modes are protected by $\hat{\mathcal{M}}_{z}$.
In addition, since $\mathcal{H}_{\mathrm{BdG}}(\boldsymbol{k})$ is $\sigma_{z}$-conserved,
the two species of helical hinge modes correspond to eigenvalues $\sigma_{z}=+1$
and $\sigma_{z}=-1$, respectively. Thus, we can consider a $\sigma_{z}$-non-coserved
term $\mathcal{H}_{\mathrm{SOC}}(\boldsymbol{k})$ which couples the helical hinge
modes by breaking $\hat{\mathcal{M}}_{z}$ symmetry and changes sign between adjacent
hinges to induce Majorana zero corner modes. The shematic diagram is plotted in Fig.\ref{Fig4}(a). 

Alternatively, exchanging the order of application of $\mathcal{H}_{\mathrm{Zeeman}}$
and $\mathcal{H}_{\mathrm{SOC}}(\boldsymbol{k})$ still achieve the same result {[}Fig.\ref{Fig4}(b){]}.
The $\hat{\mathcal{M}}_{z}$ symmetry is broken with $\mathcal{H}_{\mathrm{SOC}}(\boldsymbol{k})$
introduced while $\hat{\mathcal{T}}$ symmetry also retains since $\hat{\mathcal{T}}^{-1}\mathcal{H}_{\mathrm{BdG}}^{\prime}(\boldsymbol{k})\hat{\mathcal{T}}=\hat{\mathcal{T}}^{-1}[\mathcal{H}_{\mathrm{TSC}}^{\mathrm{total}}(\boldsymbol{k})+\mathcal{H}_{\mathrm{SOC}}(\boldsymbol{k})]\hat{\mathcal{T}}=\mathcal{H}_{\mathrm{BdG}}^{\prime}(-\boldsymbol{k})$.
And the sign-changing of $\mathrm{M}_{\phi}$ enforces two pairs of helical hinge
modes which are protected by $\hat{\mathcal{T}}$. Thus we can gap out the gapless
hinge modes by considering $\hat{\mathcal{T}}$-breaking Zeeman term $\mathcal{H}_{\mathrm{Zeeman}}$,
leading to Majorana zero corner modes. We emphasize that the conclusion based on
symmetry analysis is consistent with that from $\tilde{\theta}$ characterization. 

\subsubsection{Physical realization}

In this subsection, we provide some physical details for the feasible realization
scheme of $\mathcal{H}_{\mathrm{TOTSC}}(\boldsymbol{k})$. Obviously, Zeeman term
$\mathcal{H}_{\mathrm{Zeeman}}=\boldsymbol{B}_{\parallel}\cdot\boldsymbol{s}$ can
be achieved by applying an in-plane magnetic field to the 3D total TSC. For the time-reversal
invariant SOC $\mathcal{H}_{\mathrm{SOC}}(\boldsymbol{k})$, we can realize this
term in a multi-orbital square lattice {[}Fig.3(a) in main text{]} with two orbitals
$\{\alpha,\beta\}$ within a site. And we find the corresponding SOC term in real
space takes the form of 
\begin{align}
\hat{H}_{\mathrm{SOC}}^{\mathrm{I}}= & \sum_{\boldsymbol{i}}-\mathrm{i}\lambda_{so,0}(\hat{C}_{\boldsymbol{i},\alpha}^{\dagger}s_{z}\hat{C}_{\boldsymbol{i},\beta}-\hat{C}_{\boldsymbol{i},\beta}^{\dagger}s_{z}\hat{C}_{\boldsymbol{i},\alpha})+\sum_{\langle\boldsymbol{i},\boldsymbol{j}\rangle}(\frac{\mathrm{i}\lambda_{so,1}}{2}\hat{C}_{\boldsymbol{i},\alpha}^{\dagger}s_{z}\hat{C}_{\boldsymbol{j},\beta}+\mathrm{h.c.}),\label{eq:25}
\end{align}
where $\hat{c}_{\boldsymbol{i},\alpha(\beta)}^{\dagger}=(c_{\boldsymbol{i},\alpha(\beta),\uparrow}^{\dagger},c_{\boldsymbol{i},\alpha(\beta),\downarrow}^{\dagger})$
and $\hat{c}_{\boldsymbol{i},\alpha(\beta)}=(c_{\boldsymbol{i},\alpha(\beta),\uparrow},c_{\boldsymbol{i},\alpha(\beta),\downarrow})$
represent the creation and annihilation operators, respectively. Here we emphasize
that the time-reversal invariant $\hat{H}_{\mathrm{SOC}}^{\mathrm{I}}$ involves
two parts: the on-site term (the former) and the nearest-neighbour term (the latter)
{[}Fig.3(b) in main text{]} with $\lambda_{so,0}$, $\lambda_{so,1}$ denoting the
corresponding SOC strengths. Notably, these orbitals $\{\alpha,\beta\}$ can be chosen
as ($p_{x},p_{y}$), ($d_{xz},d_{yz}$), ($d_{xy},d_{x^{2}-y^{2}}$). Remarkly, the
realistic superconducting system $\beta$-$\mathrm{PdBi_{2}}$ which exhibits $p$-wave
pairing and topological surface states meets this condition since the maximal contributions
to its density of states at the Fermi level come from $\mathrm{Bi}$ $6p_{x+y}$
orbitals, indicating that it may serve as a potential candidate.

\subsection{\label{subsec:Realization scheme for TOTIs}Realization scheme for TOTIs }

\begin{figure*}
\includegraphics[scale=1.1]{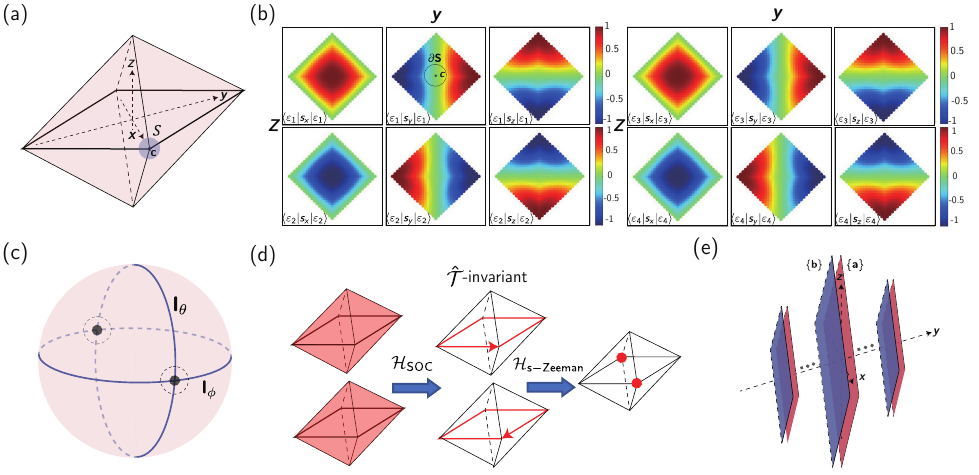}

\caption{\label{Fig5}(a) The sample geometry. We consider a regular octahedron here and focus
on an area $\mathcal{\boldsymbol{S}}$ centered at corner $\mathrm{\boldsymbol{c}}$.
(b) The distributions of surface (pseudo)spins ($\mathrm{\boldsymbol{n}}_{1},\mathrm{\boldsymbol{n}}_{2},\mathrm{\boldsymbol{n}}_{3},\mathrm{\boldsymbol{n}}_{4}$)
expectations on the $y-z$ plane. The black circle denotes $\mathcal{\partial\mathcal{\boldsymbol{S}}}$
and $\mathrm{\boldsymbol{c}}$ is target corner. (c) $\tilde{\theta}$ characterization
for TOTIs. $\mathrm{\boldsymbol{l}}_{\phi}$ and $\mathrm{\boldsymbol{l}}_{\theta}$
represent the mass domain walls generated by $\mathcal{H}_{\mathrm{SOC}}$ and $\mathcal{H}_{\mathrm{s-Zeeman}}$,
respectively. The black dots, which are intersections of $\mathrm{\boldsymbol{l}}_{\phi}$
and $\mathrm{\boldsymbol{l}}_{\theta}$, stand for surface mass field vortices and
contribute to nonzero winding numbers. (d) Symmetry analysis for the emergence of
zero corner modes. (e) The arrangement of sublattice layers $\{a\}$ and $\{b\}$.}
\end{figure*}

The proposal for TOTIs exploits the stagger Zeeman term and SOC term with model Hamiltonian
given by 
\begin{equation}
\mathcal{H}_{\mathrm{TOTIs}}(\boldsymbol{k})=\mathcal{H}_{\mathrm{TI}}^{\mathrm{total}}(\boldsymbol{k})+\mathcal{H}_{\mathrm{s-Zeeman}}+\mathcal{H}_{\mathrm{SOC}}.
\end{equation}
The Hamiltonian for total 3D TI ($\mathcal{H}_{\mathrm{TI}}^{a}\oplus\mathcal{H}_{\mathrm{TI}}^{b}$)
reads $\mathcal{H}_{\mathrm{TI}}^{\mathrm{total}}(\boldsymbol{k})=(m-t\sum_{i}\cos k_{i})\rho_{0}s_{0}\sigma_{z}+t(\sum_{i}\sin k_{i}\rho_{0}s_{i})\sigma_{x}$,
where $\{\boldsymbol{\rho}\}$,$\{\boldsymbol{\sigma}\}$ and $\{\boldsymbol{s}\}$
are Pauli matrices in sublattice, orbital and spin space, respectively. In addition,
the stagger Zeeman term reads $\mathcal{H}_{\mathrm{s-Zeeman}}=\rho_{z}\mathrm{\boldsymbol{B}_{\parallel}}\cdot\boldsymbol{s}$
and SOC term considered here still keeps time-reversal symmetry as $\hat{\mathcal{T}}^{-1}\mathcal{H}_{\mathrm{SOC}}\hat{\mathcal{T}}=\hat{\mathcal{T}}^{-1}(\rho_{y}\mathrm{\boldsymbol{M}}\cdot\boldsymbol{s})\hat{\mathcal{T}}=\hat{\mathcal{T}}^{-1}(\lambda_{so}\rho_{y}s_{z})\hat{\mathcal{T}}=\lambda_{so}\rho_{y}s_{z}=\mathcal{H}_{\mathrm{SOC}}$. 

\subsubsection{$\tilde{\mathrm{n}}_{\mathrm{WZ}}$ mechanism }

We first solve the topological surface states on surface $\boldsymbol{\Sigma}(\theta,\varphi)$:
\begin{align}
|\varepsilon_{1}\rangle=\left(\begin{array}{c}
1\\
0
\end{array}\right)_{\rho}\otimes\frac{1}{\sqrt{2}}\left(\begin{array}{c}
1\\
-i
\end{array}\right)_{\sigma}\otimes\left(\begin{array}{c}
\cos\theta/2\\
e^{i\phi}\sin\theta/2
\end{array}\right)_{s};\quad & |\varepsilon_{2}\rangle=\left(\begin{array}{c}
1\\
0
\end{array}\right)_{\rho}\otimes\frac{1}{\sqrt{2}}\left(\begin{array}{c}
1\\
i
\end{array}\right)_{\sigma}\otimes\left(\begin{array}{c}
\sin\theta/2\\
-e^{i\phi}\cos\theta/2
\end{array}\right)_{s}\nonumber \\
|\varepsilon_{3}\rangle=\left(\begin{array}{c}
0\\
1
\end{array}\right)_{\rho}\otimes\frac{1}{\sqrt{2}}\left(\begin{array}{c}
1\\
-i
\end{array}\right)_{\sigma}\otimes\left(\begin{array}{c}
\cos\theta/2\\
e^{i\phi}\sin\theta/2
\end{array}\right)_{s};\quad & |\varepsilon_{4}\rangle=\left(\begin{array}{c}
0\\
1
\end{array}\right)_{\rho}\otimes\frac{1}{\sqrt{2}}\left(\begin{array}{c}
1\\
i
\end{array}\right)_{\sigma}\otimes\left(\begin{array}{c}
\sin\theta/2\\
-e^{i\phi}\cos\theta/2
\end{array}\right)_{s}.\label{eq:27}
\end{align}
Projecting the two constant terms $(\rho_{z}\mathrm{\boldsymbol{B}_{\parallel}}\cdot\boldsymbol{s},\rho_{y}\mathrm{\boldsymbol{M}}\cdot\boldsymbol{s})$
onto surface states subspace yields a specialized surface mass field $(\boldsymbol{e}_{r}\cdot\mathrm{\boldsymbol{B}_{\parallel}},\boldsymbol{e}_{r}\cdot\mathrm{\boldsymbol{M}})$.
The (pseudo)spin polarizations are $\{\mathrm{\boldsymbol{n}}_{1},\mathrm{\boldsymbol{n}}_{2}\}_{a}$
and $\{\mathrm{\boldsymbol{n}}_{3},\mathrm{\boldsymbol{n}}_{4}\}_{b}$ where $\mathrm{\boldsymbol{n}}_{i}=\langle\varepsilon_{i}|\boldsymbol{s}|\varepsilon_{i}\rangle$
with $\mathrm{\boldsymbol{n}}_{1}=-\mathrm{\boldsymbol{n}}_{2}=\mathrm{\boldsymbol{n}}_{3}=-\mathrm{\boldsymbol{n}}_{4}=\boldsymbol{e}_{r}$.
As shown in Fig.\ref{Fig5}(a), we focus on a regular octahedron sample and take
target corner $\mathrm{\boldsymbol{c}}$ as an example to predict the emergence of
zero corner modes in terms of $\tilde{\mathrm{n}}_{\mathrm{WZ}}$ mechanism. Now
let's consider an area $\mathcal{\boldsymbol{S}}$ and study the surface (pseudo)spin
textures on it. As we can see, these textures {[}Fig.\ref{Fig5}(b) and Fig.\ref{Fig4}(c){]}
can't constitute the perfect unit spheres, so we need ``continuation'' {[}Fig.4(c)
in main text{]} to complement the real (pseudo)spin textures. Moreover, we find $\boldsymbol{e}_{\boldsymbol{\mathrm{M}}\times\mathrm{\boldsymbol{B}_{\parallel}}}=\boldsymbol{e}_{r=r_{\mathrm{\boldsymbol{c}}}}(\boldsymbol{e}_{x})$
since $\mathrm{\boldsymbol{B}_{\parallel}}=(0,-1,0)$ and $\mathrm{\boldsymbol{M}}=(0,0,1)$.
Therefore, we can get $\tilde{\mathrm{n}}_{\mathrm{WZ}}^{a}=\tilde{\mathrm{n}}_{\mathrm{WZ}}^{b}=1$
and predict the existence of zero corner mode with $(\rho_{z}\mathrm{\boldsymbol{B}_{\parallel}}\cdot\boldsymbol{s},\rho_{y}\mathrm{\boldsymbol{M}}\cdot\boldsymbol{s})$
introduced. This TOTI totally supports two zero corner modes and numerical results
shown in Fig.4(d) in main text favor such conclusions. 

\subsubsection{Symmetry analysis }

The symmetry analysis can also guide us to identify zero corner mode. As shown in
Fig.\ref{Fig5}(d), the 3D total TI $\mathcal{H}_{\mathrm{TI}}^{\mathrm{total}}$
involves two copies of minimal TIs $\mathcal{H}_{\mathrm{TI}}^{a}(\boldsymbol{k})$
and $\mathcal{H}_{\mathrm{TI}}^{b}(\boldsymbol{k})$ which are characterized by the
same $\mathbb{Z}_{2}$ index ($\mathbb{Z}_{2}=1$). Thus the surface Dirac cones
of total TI can be gapped out by effective mass $\mathrm{\tilde{M}}_{\phi}$ derived
from time-reversal invariant SOC term $\mathcal{H}_{\mathrm{SOC}}=\rho_{y}\mathrm{\boldsymbol{M}}\cdot\boldsymbol{s}$
and the sign-inversion of $\mathrm{\tilde{M}}_{\phi}$ between adjacent surfaces
leads to helical hinge modes protected by $\hat{\mathcal{T}}$-symmetry. In order
to obtain the zero corner modes, $\mathcal{\hat{T}}$-breaking in-plane staggered
Zeeman field $\mathcal{H}_{\mathrm{s-Zeeman}}=\rho_{z}\mathrm{\boldsymbol{B}_{\parallel}}\cdot\boldsymbol{s}$
is necessary. Notice that the Hamiltonian $\mathcal{H}_{\mathrm{TOTIs}}(\boldsymbol{k})$
considered here does not maintain inversion symmetry $\mathcal{\hat{I}}=\rho_{z}\sigma_{z}$
since $\mathcal{\hat{I}}^{-1}\mathcal{H}_{\mathrm{TOTIs}}(\boldsymbol{k})\mathcal{\hat{I}}\neq\mathcal{H}_{\mathrm{TOTIs}}(-\boldsymbol{k})$,
which agrees with the classification theory proposed in \citep{KhalafPRB2018-1}. 

\subsubsection{Physical realization}

As displayed in Fig.\ref{Fig5}(e), the bipartite square lattices are adopted and
$(a,b)$ denote two different sublattice sites within a unit cell. Moreover, the
2D layers $\{a\}$ and $\{b\}$ are parallel to $x-y$ plane and arranged alternatively
along $y$-axis. Therefore, the staggered Zeeman term $\mathcal{H}_{\mathrm{s-Zeeman}}=\rho_{z}\boldsymbol{B}_{\parallel}\cdot\boldsymbol{s}$
with $\mathrm{\boldsymbol{B}_{\parallel}}=(0,-1,0)$ can be achieved by applying
antiferromagnetism between layer $\{a\}$ and $\{b\}$. For $\mathcal{\hat{T}}$-invariant
SOC term considered here, we find it only remains the real space intra-cell term
\begin{equation}
\hat{H}_{\mathrm{SOC}}^{\mathrm{II}}=\sum_{\boldsymbol{i}}-\mathrm{i}\lambda_{so}(\hat{C}_{\boldsymbol{i},a}^{\dagger}s_{z}\hat{C}_{\boldsymbol{i},b}-\hat{C}_{\boldsymbol{i},b}^{\dagger}s_{z}\hat{C}_{\boldsymbol{i},a}).\label{eq:28}
\end{equation}

In the following, we take the 3D bulk $\mathrm{MnBi_{2}Te_{4}}$ which is predicted
as the antiferromagnetic (AFM) TI as a example to demonstrate the realization of
TOTI phase. $\mathrm{MnBi_{2}Te_{4}}$ is a tetradymite compound consisting of stacked
Te-Bi-Te-Mn-Te-Bi-Te septuple layers (SLs). And the spins of Mn exhibit the intralayer
ferromagnetism (within a single SL) and interlayer antiferromagnetism (between neighbouring
SLs) . Here we consider the sample geometry shown in Fig.\ref{Fig5}(a) and AFM orders
along $y$-direction. The staggered Zeeman term $\mathcal{H}_{\mathrm{s-Zeeman}}$
can be provided by the interlayer antiferromagnetism. Thus, a pair of helical edge
states emerge at corresponding hinges. In addition, the $\mathcal{\hat{T}}$-invariant
SOC term $\hat{H}_{\mathrm{SOC}}^{\mathrm{II}}$ is promising to be realized by introduce
the spin-dependent couplings between Mn atoms of neighbouring SLs, which couples
the helical edge states and contributes a sign-inversion mass of adjacent hinges.
As a result, we can obtain the 3rd-order corner modes. According to the above discussion,
we conjecture that the 3D AFM TI $\mathrm{MnBi_{2}Te_{4}}$ may be a ideal platform
to realize the TOTIs.

\end{document}